\documentclass[a4paper,12pt]{article}
\pdfoutput=1 % if your are submitting a pdflatex (i.e. if you have
\usepackage{jheppub} % for details on the use of the package, please
\usepackage[T1]{fontenc} % if needed

\usepackage{amssymb,amsmath,amsbsy}
\usepackage{booktabs}
\usepackage{float}
\usepackage{multirow}
\usepackage{graphicx}
\usepackage{placeins}

\def\ti              {\tilde}
\def\nt              {\ti\chi^0}
\def\st              {\ti t}

\def\PL              {P_L^{}}
\def\PR              {P_R^{}}

\def\ETmiss {E_T^{\rm miss}}
\newcommand{\Lag}{\mathcal L}
\newcommand{\MET}{E_T^{\rm miss}}

\newcommand{\sfrac}[2] {{\textstyle \frac{#1}{#2}}}

\title{\boldmath  Scalar versus fermionic top partner interpretations of $t\bar t + E_T^{\rm miss}$ searches at the LHC}
\author[a]{Sabine Kraml,}
\author[a,b]{Ursula Laa,}
\author[c,d]{Luca Panizzi,}
\author[c,d]{Hugo Prager}

\affiliation[a]{Laboratoire de Physique Subatomique et de Cosmologie, Universit\'e Grenoble-Alpes,
CNRS/IN2P3, 53 Avenue des Martyrs, F-38026 Grenoble, France}
\affiliation[b]{LAPTH, Universit\'e Savoie Mont Blanc, CNRS, B.P.~110, F-74941 Annecy-le-Vieux, France}
\affiliation[c]{School of Physics and Astronomy, University of Southampton, Highfield, Southampton SO17 1BJ, UK}
\affiliation[d]{Particle Physics Department, Rutherford Appleton Laboratory, Chilton, Didcot, Oxon OX11 0QX, UK}

\emailAdd{sabine.kraml@lpsc.in2p3.fr}
\emailAdd{ursula.laa@lpsc.in2p3.fr}
\emailAdd{l.panizzi@soton.ac.uk}
\emailAdd{hugo.prager@soton.ac.uk}

\preprint{LPSC16150}

\abstract{We assess how different ATLAS and CMS searches for supersymmetry in the $t\bar t + E_T^{\rm miss}$ final state at Run~1 of the LHC constrain scenarios with a fermionic top partner and a dark matter candidate. We find that the efficiencies of these searches in all-hadronic, 1-lepton and 2-lepton channels are quite similar for scalar and fermionic top partners. Therefore, in general, efficiency maps for stop--neutralino simplified models can also be applied to fermionic top-partner models, provided the narrow width approximation holds in the latter. Owing to the much higher production cross-sections of heavy top quarks as compared to stops, masses up to $m_T\approx 850$~GeV can be excluded from the Run~1 stop searches.  
Since the simplified-model results published by ATLAS and CMS do not extend to such high masses, we provide our own efficiency maps obtained with {\sc CheckMATE} and {\sc MadAnalysis\,5} for these searches. 
Finally, we also discuss how generic gluino/squark searches in multi-jet final states constrain heavy top partner production.}

\begin{document} 
\maketitle
\flushbottom

\newpage
%==============================================================================
\section{Introduction}
\label{sec:intro}
%==============================================================================

After the discovery of the Higgs boson \cite{Aad:2012tfa, Chatrchyan:2012xdj}, the quest for new physics beyond 
the Standard Model (SM) is arguably the most pressing open issue in particle physics. 
If this new physics is responsible for the dark matter (DM) of the universe in the form of weakly interacting massive particles, its signatures at the LHC and other future colliders are expected to be characterized by events with an excess of missing transverse energy, $\MET$. An intense experimental effort is thus being made at the LHC to isolate such signatures, though no signal has been observed so far.\footnote{Of course, $\MET$ signatures cannot be univocally associated with the production of DM. Neutral long-lived particles which decay outside the detector would produce the very same signatures without being DM. However, the observation of a signature compatible with DM at the LHC would allow to focus on specific regions of the parameter space to be corroborated by other observations, like DM direct and/or indirect detection.}

The prototype for a new physics model leading to $\MET$ signatures is R-parity conserving supersymmetry (SUSY), 
in particular the minimal supersymmetric standard model (MSSM) with a neutralino as the lightest supersymmetric 
particle (LSP) \cite{Martin:1997ns,Drees:2004jm,Baer:2006rs}.  
Indeed, a large number of searches for final states containing jets and/or leptons plus $\MET$ have been designed by the ATLAS and CMS SUSY groups~\cite{ATLAS:twiki,CMS:twiki}, and the interpretations of the results  
are typically limits in some SUSY simplified model. 
Examples are multi-jet + $\MET$ searches being interpreted as limits in the the gluino--neutralino mass plane, 
or searches for the $t\bar t+\MET$ final state being interpreted in terms of stops decaying to top+neutralino. 

The same searches can be used to put constraints on scenarios leading to final states with $\MET$ generated by the production of extra quarks (XQs) decaying to a bosonic DM candidate. This occurs for instance in Universal Extra Dimensions (UED)~\cite{Antoniadis:1990ew,Appelquist:2000nn,Servant:2002aq,Csaki:2003sh,Cacciapaglia:2009pa}, 
Little Higgs models with T-parity~\cite{Cheng:2003ju,Cheng:2004yc,Low:2004xc,Hubisz:2004ft,Hubisz:2005tx,Cheng:2005as}, 
or generically any model with extra matter and a $\mathcal Z_2$ parity under which the SM particles are even and (part of) the new states are odd. A common feature of these models is that the new states have the same spin as their SM partners, while in SUSY the spins differ by half a unit.

In all these models, the lightest odd particle is a DM candidate which interacts with the SM states through new mediator particles. A crucial property of scenarios where the mediators are odd is that they can only be produced in pairs or in association with other odd particles. This is then followed by (cascade) decays into SM particles and the DM candidate. Since the spins in the decays are all correlated, if it was possible to identify the spin of the mediator, this would give information on the bosonic/fermionic nature of the DM candidate as well.

It is therefore interesting to ask how the current results from SUSY searches constrain other models of new physics that would lead to the same signatures, and how same-spin and different-spin scenarios could be distinguished should a signal be observed. In this paper,  
we concentrate on the first of these questions, comparing the cases 
of pair production of scalar (SUSY) and 
fermionic (XQ) top partners with charge 2/3, which decay into $t+{\rm DM}$,\footnote{Here and in the following, we understand ``DM'' as the dark matter {\em candidate}, i.e.\ a neutral massive particle that escapes detection as $\MET$ but whose astrophysical properties remain open.} 
thus leading to a $t\bar t + \MET$ final state. Concretely, we consider the processes 
\begin{eqnarray*}
 \text{Top partner with spin 0: } &\quad& pp \to \tilde t\, \tilde t^* \to t \bar t + \tilde\chi^0 \tilde\chi^0 \\
 \text{Top partner with spin 1/2: } &\quad& pp \to T \bar T \to t \bar t + \{S^0 S^0 \text{ or } V^0 V^0\}
\end{eqnarray*}
where $\tilde\chi^0$, $S^0$ and $V^0$ represent fermionic, scalar, and vectorial DM candidates respectively. 
Recasting a number of ATLAS and CMS searches for stops~\cite{ATLAS:2013cma,Aad:2014kra,Chatrchyan:2013xna,Aad:2014qaa} from Run~1 of the LHC, as well as a generic search for gluinos and squarks~\cite{Aad:2014wea} by means 
of {\sc CheckMATE}~\cite{Drees:2013wra} and {\sc MadAnalysis}\,5~\cite{Conte:2014zja,Dumont:2014tja},  
we compare the efficiencies of these searches for the processes above. 
This allows us to determine whether cross-section upper limit maps or efficiency maps derived in the context of stop--neutralino  simplified models can safely be applied to XQ scenarios where the $t\bar t + \MET$ final state arises from the production of heavy $T$ quarks. Such maps are used in public tools like {\sc SModelS}~\cite{Kraml:2013mwa,Kraml:2014sna} 
and {\sc XQCAT}~\cite{Barducci:2014ila,Barducci:2014gna}, and it is relevant to know how generically they can be applied. 
Moreover, we determine up-to-date bounds in the parameter space of the XQ and DM masses -- 
such bounds were posed by a few early searches at the Tevatron \cite{Aaltonen:2011rr, Aaltonen:2011na} 
and the LHC at 7~TeV \cite{ATLAS:2011mda, CMS:2012dwa}, but can be improved by a reinterpretation of the 8~TeV LHC results as we do in this paper.

Related studies exist in the literature. In particular, a re-interpretation of a few ATLAS and CMS SUSY searches at 7~TeV in terms of UED signatures was done in \cite{Cacciapaglia:2013wha}, using among others a simplified scenario with top-partners decaying to DM and light quarks. 
The applicability of SUSY simplified model results to new physics scenarios with same-spin SM partners was analysed in \cite{Edelhauser:2015ksa} also in the context of UED, focussing on the so-called T2 topology which corresponds to squark-antisquark production in the limit of a heavy gluino.    
The effect of a different spin structure for the $l^+l^- + \MET$ final state was studied in~\cite{Arina:2015uea}. Recently, a study of constraints and LHC signatures of a scenario with a vector-like top partner decaying to a top quark and scalar DM has been performed in~\cite{Baek:2016lnv}. 
Here, we extend these works by considering specifically top partners and by applying up-to-date recasting tools. 

The structure of the paper is the following. In Section~\ref{sec:models}, we describe the simplified models we use for the SUSY and XQ scenarios and define the benchmark points we consider for our analysis. 
The tools we use and the processes we consider are described in Section~\ref{sec:MC}, together with selected kinematical distributions at generator level which are useful for a better understanding of our results. 
Section~\ref{sec:analyses8tev} provides detailed descriptions of the experimental analyses and the effects found for our benchmark points. The results are then summarized in the top-partner versus DM mass plane in Section~\ref{sec:bounds8TeV}.
Section~\ref{sec:conclusions} contains our conclusions. 
A few additional results and comparisons which may be interesting to the reader are presented in Appendix~A. 
The event numbers from the experimental analyses are listed in Appendix~B.

%==============================================================================
\section{Benchmark scenarios}
\label{sec:models}
%==============================================================================

%-----------------------------------------------------------------------------------------------------------------------------------------
\subsection{The SUSY case: stop--neutralino simplified model}
%-----------------------------------------------------------------------------------------------------------------------------------------

The prototype for the $t\bar t + E_T^{\rm miss}$ signature in the SUSY context is a stop--neutralino simplified model. 
This assumes that the lighter stop, $\tilde t_1$, and the lightest neutralino, $\tilde\chi^0_1$, taken to be the lightest SUSY particle and the DM candidate, are the only accessible sparticles --- all other sparticles are assumed to be heavy. 
In this case, direct stop pair production is the only relevant SUSY production mechanism. Moreover, for large enough mass difference, the $\tilde t_1$ decays to 100\% into $t+\tilde\chi^0_1$. The process we consider thus is 
\begin{equation}
  pp\to \tilde t_1^{} \tilde t_1^* \to t\bar t \tilde\chi^0_1\tilde\chi^0_1 \,.
\end{equation}
Following the notation of \cite{Gajdosik:2004ed}, the top--stop--neutralino interaction is given by  ($i=1,2$; $k=1,...,4$)
\begin{align}
  {\cal L}_{t\st\nt}  
   &= g\,\bar t\,( f_{Lk}^{\st}\PR + h_{Lk}^{\st}\PL )\,\nt_k\,\st_L^{} +
      g\,\bar t\,( h_{Rk}^{\st}\PR + f_{Rk}^{\st}\PL )\,\nt_k\,\st_R^{} + {\rm h.c.} \nonumber \\
   & = g\,\bar t\,( a^{\,\st}_{ik}\PR + b^{\,\st}_{ik}\PL )\,\nt_k\,\st_i^{} + {\rm h.c.}
  \label{eq:LagSUSYSMS}
\end{align}
where $P_{R,L}^{}  = \frac{1}{2}(1\pm\gamma_5)$ are the right and left projection operators,  and 
\begin{align}
   a^{\,\st}_{ik} &= f_{Lk}^{\st}\,R_{i1}^{\st} +
                      h_{Rk}^{\st}\,R_{i2}^{\st}\,,  \nonumber \\ % \label{eq:aik}\\
   b^{\,\st}_{ik} &= h_{Lk}^{\st}\,R_{i1}^{\st} +
                      f_{Rk}^{\st}\,R_{i2}^{\st}\,.   \label{eq:bik}
\end{align}
The $f_{L,R}^{\st}$ and $h_{L,R}^{\st}$ couplings are
\begin{align}
  f_{Lk}^{\,\ti t} &= -\sfrac{1}{\sqrt 2}\,(N_{k2} +\sfrac{1}{3}\tan\theta_W N_{k1}) \,, \nonumber\\
  f_{Rk}^{\,\ti t} & = \sfrac{2\sqrt 2}{3}\,\tan\theta_W N_{k1}\,, \hspace{24mm} % \nonumber\\
  h_{Rk}^{\ti t} = -y_t\, N_{k4} = h_{Lk}^{\ti t*}\,, 
\end{align}
with $N$ the neutralino mixing matrix and $y_t=m_t/(\sqrt{2}m_W\sin\beta)$ the top Yukawa coupling in the MSSM. 
Finally, $R$ is the stop mixing matrix,  
\begin{equation}
  {\tilde t_1 \choose \tilde t_2}  = R \, {\tilde t_L \choose \tilde t_R} \,, \quad 
  R =  \left(\begin{array}{rr} \cos\theta_{\tilde t} & \sin\theta_{\tilde t} \\ -\sin\theta_{\tilde t} & \cos\theta_{\tilde t} \end{array}\right) \,.
\end{equation}
All this follows SLHA \cite{Skands:2003cj} conventions. 

Under the above assumption that all other neutralinos besides the $\tilde\chi^0_1$ and the charginos are heavy, the $\tilde\chi^0_1$ is dominantly a bino.  Neglecting the wino and higgsino components $N_{12}$ and $N_{14}$, the 
$t\st_1\nt_1$ interaction from Eq.~\eqref{eq:LagSUSYSMS}  simplifies to 
\begin{align}
  {\cal L}_{t\st_1\nt_1}  
  & \approx - \frac{g}{3\sqrt{2}} \tan\theta_W N_{11} \, \bar t 
   \left( \cos\theta_{\tilde t}\, P_R^{} - 4 \sin\theta_{\tilde t}\, P_L^{} \right) \tilde\chi^0_1\, \tilde t_1^{} + {\rm h.c.}\,.
  \label{eq:LagSUSYSMSsimple}
\end{align}
While in practice one never has a {\em pure} bino,  this approximation shows that the polarisation of the  
tops originating from the $\tilde t_1\to t\tilde\chi^0_1$ decays will reflect the chirality of the $\tilde t_1$.
(The wino interaction also preserves the chirality, while the higgsino one flips it.)
This will be relevant for defining XQ benchmark scenarios analogous to SUSY ones, since the $p_T$ and angular distributions of the top decay products somewhat depend on the top polarisation~\cite{Cao:2006wk,Nojiri:2008ir,Shelton:2008nq,Perelstein:2008zt,Berger:2012an,Chen:2012uw,Bhattacherjee:2012ir,Belanger:2012tm,Low:2013aza,Belanger:2013gha,Wang:2015ola}.

%-----------------------------------------------------------------------------------------------------------------------------------------
\subsection{The extra quark scenario: conventions and Lagrangian terms}
%-----------------------------------------------------------------------------------------------------------------------------------------

As the XQ analogue of the SUSY case above, we consider a minimal extension of the SM with one extra quark state and one DM state, assuming that the XQ mediates the interaction between the DM and the SM quarks of the third generation. Interactions between the XQ, DM and lighter quarks are neglected. 
The most general Lagrangian terms depend on the representation of the DM and of the XQ. We label XQ singlet states as $T$ (with charge $+2/3$) or $B$ (with charge $-1/3$) and XQ doublet states as $\Psi_Y$, where $Y$ corresponds to the weak hypercharge of the doublet in the convention $Q=T_3+Y$, with $Q$ the electric charge and $T_3$ the weak isospin. The doublets can then be $\Psi_{1/6}={T \choose B}$ or states which contain exotic components $\Psi_{7/6}={X_{5/3} \choose T}$ and $\Psi_{-5/6}={B \choose Y_{-4/3}}$. The DM states are labelled as $S^0_{\rm DM}$ if scalar singlets or $V^{0\mu}_{\rm DM}$ if vector singlets; if the DM belongs to a doublet representation, the multiplet is labelled as $\Sigma_{\rm DM}={S^+ \choose S^0_{\rm DM}}$ (with the charge conjugate $\Sigma^c_{\rm DM}={S^{0}_{\rm DM} \choose -S^-}$) if scalar or $\mathcal{V}_{\rm DM}={V^+ \choose V^0_{\rm DM}}$ (with the charge conjugate $\mathcal{V}^c_{\rm DM}={V^0_{\rm DM} \choose V^-_{\rm DM}}$) if vector. 
The couplings between the XQ, the DM and the SM quarks are denoted as $\lambda_{ij}^q$ if the DM is scalar, or $g_{ij}^q$ if the DM is vector: the labels $\{i,j\}=1,2$ indicate the representations of the XQ and DM respectively (1 for singlet, 2 for doublet), while $q=t,b$ identifies which SM quark the new states are coupled with, in case of ambiguity. 
We classify below the Lagrangian terms for the minimal SM extensions with one XQ and one DM representation (singlets and doublets) but we anticipate that in the following, for simplicity, we will only consider scenarios with a DM singlet. 
\begin{itemize}
\item Lagrangian terms for a \textit{DM singlet}. A DM singlet can couple either with a XQ singlet or with a XQ doublet $\Psi_{1/6}={T \choose B}$.
\begin{eqnarray}
\Lag^S_1 &=& 
\left[
\lambda_{11}^t \bar{T} P_R^{} t + 
\lambda_{11}^b \bar{B} P_R^{} b +
\lambda_{21} \overline\Psi_{1/6} P_L^{} {t \choose b} 
\right] 
S^0_{\rm DM} + {\rm h.c.} 
\label{eq:LagSingletDMS}
\\
\Lag^V_1 &=& 
\left[
g_{11}^t \bar{T} \gamma_\mu P_R^{} t + 
g_{11}^b \bar{B} \gamma_\mu P_R^{} b + 
g_{21}  \overline\Psi_{1/6} \gamma_\mu P_L^{} {t \choose b} 
\right] 
V^{0\mu}_{\rm DM} + {\rm h.c.} 
\label{eq:LagSingletDMV}
\end{eqnarray}
\item Lagrangian terms for a \textit{DM doublet}. A DM doublet can couple with XQ singlets or doublets with different hypercharges.
\begin{eqnarray}
\Lag^S_2 &=& 
\left[
\lambda_{12}^b \bar B P_L^{} {t \choose b} +
\lambda_{22}^b \overline\Psi_{1/6} P_R^{} b +
(\lambda_{22}^{t})^\prime \overline\Psi_{5/6} P_R^{} t
\right] \Sigma_{\rm DM} \nonumber \\
&+&\left[
\lambda_{12}^t \bar T P_L^{} {t \choose b} +
\lambda_{22}^t \overline\Psi_{1/6} P_R^{} t +
(\lambda_{22}^{b})^\prime \overline\Psi_{-1/6} P_R^{} b 
\right] \Sigma^c_{\rm DM} \\
\Lag^V_2 &=& 
\left[
g_{12}^b \bar B \gamma_\mu P_L^{} {t \choose b} +
g_{22}^b\overline\Psi_{1/6} \gamma_\mu P_R b +
(g_{22}^{t})^\prime \overline\Psi_{5/6} \gamma_\mu P_R^{} t
\right] \mathcal{V}_{\rm DM}^\mu \nonumber \\
&+&\left[
g_{12}^t \bar T \gamma_\mu P_L^{} {t \choose b} +
g_{22}^t \overline\Psi_{1/6} \gamma_\mu P_R^{} t +
(g_{22}^{b})^\prime \overline\Psi_{-1/6} \gamma_\mu P_R^{} b 
\right] \mathcal{V}_{\rm DM}^{c,\mu} 
\label{eq:LagDoubletDM}
\end{eqnarray}
\end{itemize}
However, in scenarios with a DM doublet, there are always additional exotic states besides the XQ partners of the SM quarks and the DM state, namely charged scalars or vectors and quarks with charges $5/3$ or $4/3$. As mentioned above, in order to stick to a minimal extension of the SM containing a partner of the top quark and the DM candidate as the only new states, in the following we consider only the Lagrangian terms of Eqs.~\eqref{eq:LagSingletDMS} or \eqref{eq:LagSingletDMV}, depending on the spin of the DM. It is also worth noticing that in the considered scenarios  the XQs do not mix with SM states because they have a different quantum number under the $\mathcal Z_2$ symmetry. 
Moreover, to focus only on top partners, we set $\lambda_{11}^b=g_{11}^b=0$. 
Depending on the representation of the XQ, one can then identify some limiting cases: 
\begin{itemize}
\item \textit{Vector-like XQ (VLQ)}. If the XQ is vector-like, the left-handed and right-handed projections belong to the same $SU(2)$ representation. Therefore if the VLQ is a singlet, only couplings with SM singlets are allowed, and $\lambda_{21}=0$ or $g_{21}=0$. On the other hand, if the VLQ is a doublet, $\lambda_{11}=0$ or $g_{11}=0$.
Unlike cases where VLQs mix with the SM quarks through Yukawa couplings via the Higgs boson, couplings for the opposite chiralities are not just suppressed, they are identically zero.
The mass term for a VLQ can be written in a gauge-invariant way as:
\begin{equation}
\Lag_{\rm VLQ} = - M_{T_{\rm VLQ}} \bar T T
\label{eq:VLQmass}
\end{equation}
where $M_{T_{\rm VLQ}}$ is a new physics mass scale not necessarily related to a Higgs-like mechanism for mass generation.

\item \textit{Chiral XQ (ChQ)}. If the XQ is chiral, all the couplings of Eqs.~\eqref{eq:LagSingletDMS} or \eqref{eq:LagSingletDMV} can be allowed at the same time. ChQs can acquire mass in a gauge invariant way via the Higgs mechanism, analogously to SM quarks:
\begin{align}
\Lag_{\rm ChQ} = & - y_{\rm XQ}^B \bar \Psi_{1/6} H B - y_{\rm XQ}^T \bar \Psi_{1/6} H^c T + {\rm h.c.} \nonumber \\  
                              & \Longrightarrow\; - M_{T_{\rm ChQ}} \bar T T - M_{B_{\rm ChQ}} \bar B B
\label{eq:ChQmass}
\end{align}
where $M_{\{T,B\}_{\rm ChQ}} = y_{\rm XQ}^{\{T,B\}} v / \sqrt{2}$ and $v$ is the Higgs VEV. At this point it has to be mentioned that the contribution of the new ChQ to Higgs production and decay processes, even if different from scenarios where a 4$^\text{th}$ chiral generation mixes with the SM quarks, can be used to pose constraints on the coupling between the XQ and the Higgs boson, and as a consequence, on the maximum mass the ChQ can acquire through the Higgs mechanism. Of course, ChQs can still acquire mass by some different new physics mechanism (for example by interacting with a heavier scalar which develops a VEV). For this reason we can consider the ChQ mass as a free parameter in the following analysis.
\end{itemize}

% \clearpage
%-----------------------------------------------------------------------------------------------------------------------------------------
\subsection{Benchmark points}
%-----------------------------------------------------------------------------------------------------------------------------------------

In order to compare the XQ and SUSY scenarios, it is useful to consider benchmark points with the same top-partner and DM masses as well as the same left and right couplings (leading to $t_L$ or $t_R$ in the final state) for the two models. To this end, we start from the stop--neutralino simplified model and  choose two mass combinations: 
$(m_{\tilde t_1},\,m_{\tilde\chi^0_1})=(600,\,10)$~GeV and $(m_{\tilde t_1},\,m_{\tilde\chi^0_1})=(600,\,300)$~GeV. 
The first one is excluded by the 8~TeV searches, while the second one lies a bit outside the 8~TeV bounds~\cite{Aad:2015pfx,CMS:2014yma,Chatrchyan:2013xna,Chatrchyan:2014lfa,CMS:2014wsa}.\footnote{The $(m_{\tilde t_1},\,m_{\tilde\chi^0_1})=(600,\,300)$~GeV mass combination actually lies just on the edge of the new 13~TeV bounds presented by CMS \cite{CMS-PAS-SUS-16-007} at the Moriond 2016 conference.}
Moreover, since the searches for $\st_1\to t\nt_1$ exhibit a small dependence on the top polarisation~\cite{Aad:2014kra}, 
we consider the two cases $\tilde t_1\sim \tilde t_R$ and $\tilde t_1\sim \tilde t_L$.\footnote{Strictly speaking, because of SU(2), a $\tilde t_1\sim \tilde t_L$ should be accompanied by a $\tilde b_L$ of similar mass; with no other 2-body decay being kinematically open, the sbottom would however decay to 100\% into $b\nt_1$and thus not contribute to the $t\bar t+\MET$signature.} The results for arbitrary stop mixing (or top polarisation) will then always lie  between these two extreme cases. This leads to four benchmark scenarios, which we denote by 
\begin{equation*}
   \rm  (600,\,10)L\,;\quad (600,\,10)R \,;\quad (600,\,300)L \,; \quad (600,\,300)R \,. 
\end{equation*}
The strategy then is to use the same mass combinations $(m_T,\,m_{\rm DM})$ and left/right couplings for the XQ case. 
For XQ$+S^0_{\rm DM}$, we directly use  $\lambda_{11}^t=b_{11}^{\st}$ and $\lambda_{21}^t=a_{11}^{\st}$. 
For XQ$+V^0_{\rm DM}$, however, the width of the XQ would be too large if we were using the same parameters as in the SUSY or scalar DM case; to preserve the narrow width approximation, we therefore reduce the couplings by a factor 10, i.e.\ $g_{11}^t=b_{11}^{\st}/10$ and $g_{21}^t=a_{11}^{\st}/10$. The concrete values for the different benchmark scenarios are listed in Table~\ref{tab:BPlist}. 

\begin{table}
\scriptsize
\centering
\begin{tabular}{c| rr | rr}
\toprule
& \multicolumn{2}{c|}{\bf (600,\,10)L} & \multicolumn{2}{c}{\bf (600,\,300)L} \\
$\tilde t_1 \sim \tilde t_L$ & 
$a_{11}^{\st} = -8.3649\;10^{-2}$ & $b_{11}^{\st} = 1.5406\;10^{-3}$ & 
$a_{11}^{\st}  = -8.3638\;10^{-2}$ & $b_{11}^{\st} = 2.5811\;10^{-3}$ \\
XQ + $S^0_{\rm DM}$ & 
$\lambda_{21}^t = -8.3649\;10^{-2}$ & $\lambda_{11}^t = 1.5406\;10^{-3}$ & 
$\lambda_{21}^t = -8.3638\;10^{-2}$ & $\lambda_{11}^t = 2.5811\;10^{-3}$ \\ 
XQ + $V^0_{\rm DM}$ & 
$g_{21}^t = -8.3649\;10^{-3}$ & $g_{11}^t = 1.5406\;10^{-4}$ & 
$g_{21}^t = -8.3638\;10^{-3}$ & $g_{11}^t = 2.5811\;10^{-4}$ \\
\midrule
& \multicolumn{2}{c|}{\bf (600,\,10)R} & \multicolumn{2}{c}{\bf (600,\,300)R} \\
$\tilde t_1 \sim \tilde t_R$ &
$a_{11}^{\st} = 1.1425\;10^{-3}$ & $b_{11}^{\st} = 3.3467\;10^{-1}$ &
$a_{11}^{\st} = 2.1823\;10^{-3}$ & $b_{11}^{\st} = 3.3466\;10^{-1}$ \\
XQ + $S^0_{\rm DM}$ & 
$\lambda_{21}^t = 1.1425\;10^{-3}$ & $\lambda_{11}^t = 3.3467\;10^{-1}$ & 
$\lambda_{21}^t = 2.1823\;10^{-3}$ & $\lambda_{11}^t = 3.3466\;10^{-1}$ \\
XQ + $V^0_{\rm DM}$ & 
$g_{21}^t = 1.1425\;10^{-4}$ & $g_{11}^t = 3.3467\;10^{-2}$ & 
$g_{21}^t = 2.1823\;10^{-4}$ & $g_{11}^t = 3.3466\;10^{-2}$ \\
\bottomrule
\end{tabular}
\caption{\label{tab:BPlist} Benchmark points for the SUSY and XQ scenarios.}
\end{table}

The alert reader will notice that in Table~\ref{tab:BPlist}, although there is a strong hierarchy between the left and right couplings, both of them are non-zero. Moreover, the couplings for the (600,\,300)L case are not the same as for the (600,\,10)L case;  
the same is true for (600,\,300)R vs.\ (600,\,10)R. 
The reason for this is as follows. The pure left or pure right case, $\tilde t_1\equiv\tilde t_L$ or $\tilde t_R$, would require that the off-diagonal entry in the stop mixing matrix is exactly zero, that is $A_t\equiv\mu/\tan\beta$, where $A_t$ is the trilinear stop-Higgs coupling, $\mu$ is the higgsino mass parameter and $\tan\beta=v_2/v_1$ is the ratio of the Higgs vacuum expectation values. To avoid such tuning, and also because the $\tilde\chi^0_1$ will never be a 100\% pure bino even if the winos and higgsinos are very heavy, we refrain from using the approximation of Eq.~\eqref{eq:LagSUSYSMSsimple} with $N_{11}=1$ and $\cos\theta_{\tilde t}=1$ or $0$.  Instead, we choose the masses of the benchmark points as desired by appropriately adjusting the relevant soft terms while setting all other soft masses to 3--5 TeV. From this we then compute the stop and neutralino mixing matrices and the full $\tilde\chi^0_1\tilde t_1t$ couplings $a_{11}^{\st}$ and $b_{11}^{\st}$ of of Eq.~\eqref{eq:LagSUSYSMS}, using 
{\sc{SuSpect}}~v2.41~\cite{Djouadi:2002ze}. 
The resulting values are $N_{11}\simeq 1$, $\cos\theta_{\tilde t}\simeq 1$ (or $\sin\theta_{\tilde t}\simeq 1$) to sub-permil precision, but nonetheless this leads to a small non-zero value of the ``other'' sub-dominant coupling, and to a slight dependence on the $\tilde\chi^0_1$ mass. 
An interesting consequence is that our comparison between SUSY and XQ is effectively between SUSY and ChQ scenarios. A comparison between SUSY and VLQ scenarios would require $\tilde t_1 \equiv \tilde t_L$ or $\tilde t_1 \equiv \tilde t_R$.
Our conclusions however do not depend on this.

%==============================================================================
\section{Monte Carlo event generation}
\label{sec:MC}
%==============================================================================

%-----------------------------------------------------------------------------------------------------------------------------------------
\subsection{Setup and tools}
%-----------------------------------------------------------------------------------------------------------------------------------------

For the Monte Carlo analysis, we simulate the $2\to 6$ process 
$$pp\to t\,\bar t~{\rm DM~DM}\to (W^+b)(W^-\bar b)~{\rm DM~DM}$$ with {\sc MadGraph}\,5~\cite{Alwall:2011uj,Alwall:2014hca}, where $\rm DM$ is the neutralino in the SUSY scenario or the scalar/vector boson in the XQ scenario.
This preserves the spin correlations in the  $t\to Wb$  decay. 
Events are then passed to {\sc Pythia}\,6~\cite{Sjostrand:2006za}, which takes care of the decay $W \to 2f$ as well as hadronisation and parton showering.\footnote{In \cite{Boughezal:2013pja} it was argued that certain kinematic distributions show sizeable differences between LO and NLO, which can be ameliorated by including initial state radiation of extra jets. We tested this but did not find any relevant differences with and without simulating extra jets for the analyses we consider in this paper.  We therefore conclude that LO matrix element plus parton showering is sufficient for the scope of this study, in particular as it saves a lot of CPU time.}

For the SUSY scenarios we make use of the MSSM model file in {\sc MadGraph}, while for the XQ simulation we implemented the model in {\sc Feynrules}~\cite{Alloul:2013bka} to obtain the UFO model format to be used inside {\sc MadGraph}.  
For the PDFs we employ the cteq6l1 set~\cite{Pumplin:2002vw}. 
To analyse and compare the effects of various ATLAS and CMS 8~TeV analyses, we employ {\sc CheckMATE}~\cite{Drees:2013wra} as well as {\sc MadAnalysis}\,5~\cite{Conte:2014zja}.
Both frameworks use {\sc Delphes\,3}~\cite{deFavereau:2013fsa} for the emulation of detector effects.

\begin{figure}[t!]
\centering\includegraphics[width=0.95\textwidth]{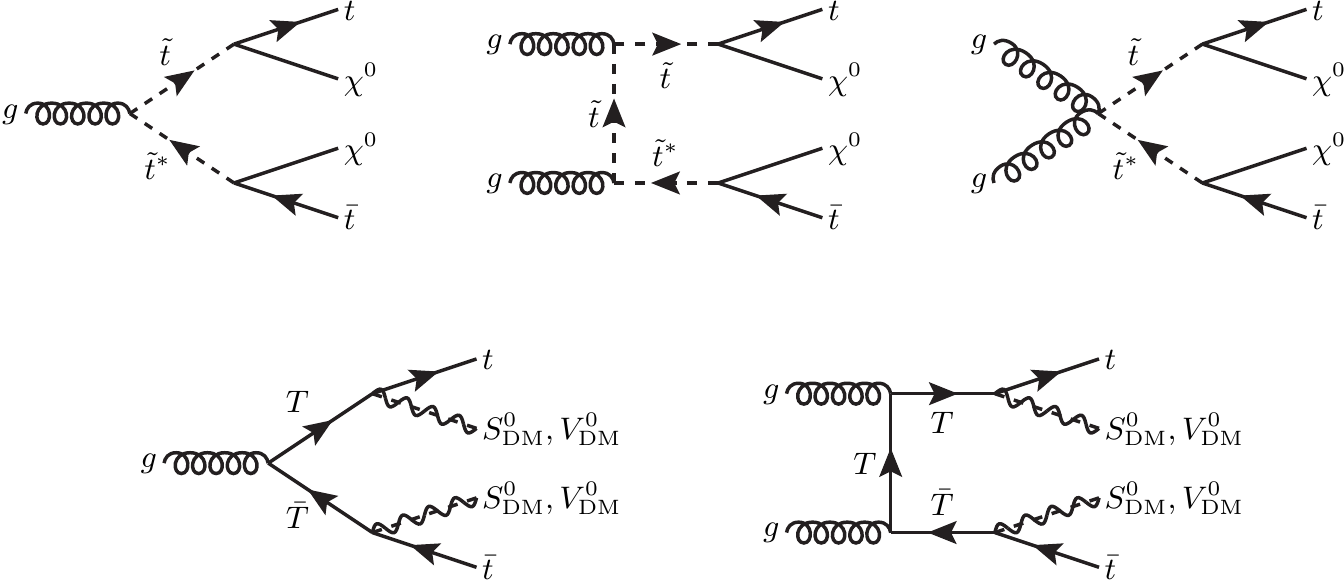}
\caption{\label{fig:topologies}Feynman diagrams for the production of $t\bar t + \MET$ in the SUSY and XQ scenarios. We have omitted for simplicity the $gg$ and $q \bar q$ initial states which are common for the s-channel gluon topologies.}
\end{figure}

\begin{figure}[t!]
\centering\includegraphics[width=0.7\textwidth]{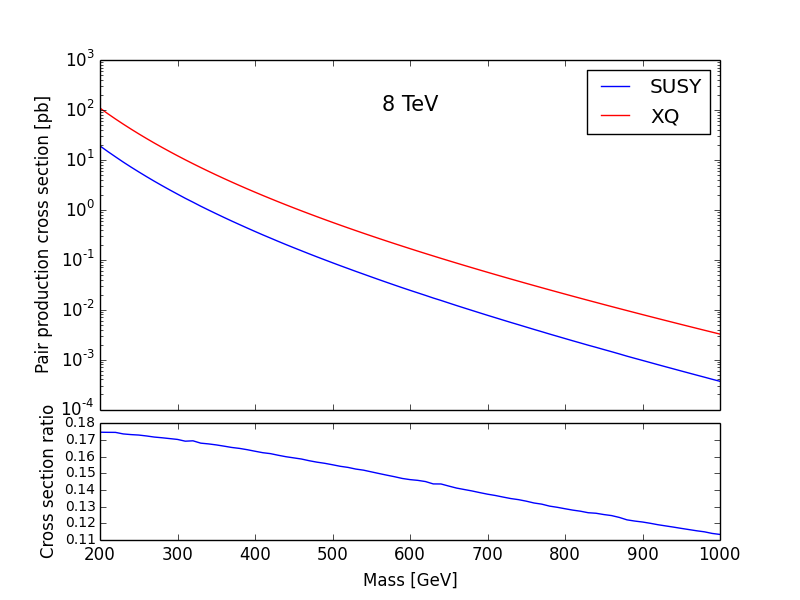} 
\caption{\label{fig:XS} Production cross-sections for SUSY and XQ top partners at $\sqrt{s}=8$~TeV. }
\end{figure}

The Feynman diagrams relevant for the SUSY and XQ processes are shown in Fig.~\ref{fig:topologies}. We observe that besides the difference in the spin of the mediator and DM, in the SUSY case there is a topology which is not present in the XQ case, namely the 4-leg diagram initiated by two gluons. 
The $pp\to \st_1^{}\st_1^*$ and $pp\to T\bar T$ production cross-sections at $\sqrt{s}=8$~TeV are compared in Fig.~\ref{fig:XS}. 
The comparison is done at the highest available order for each scenario, i.e.\ at NLO+NLL for SUSY~\cite{nllfast,Beenakker:1996ch,Kulesza:2008jb,Kulesza:2009kq,Beenakker:2009ha,Beenakker:2011fu,Beenakker:1997ut,Beenakker:2010nq} and at NLO+NNLL for XQ~\cite{Cacciari:2011hy}. We see that, for the same mass, the XQ cross-section is about a factor 5--10 larger than the SUSY cross-section.  The same experimental analysis targeting $t\bar t+\MET$ will therefore have a significantly higher reach in fermionic (XQ)  than in scalar (SUSY) top partner masses. For instance, an excluded cross-section of 20\,fb corresponds to $m_{\st_1}\gtrsim 620$~GeV in the SUSY case but  $m_{T}\gtrsim 800$~GeV in the XQ case. The precise reach will, of course,  depend on the specific cut acceptances in the different models.

%-----------------------------------------------------------------------------------------------------------------------------------------
\subsection{Generator-level distributions}
%-----------------------------------------------------------------------------------------------------------------------------------------

\begin{figure}[h!]\vspace*{4mm}\centering
\includegraphics[width=0.34\textwidth]{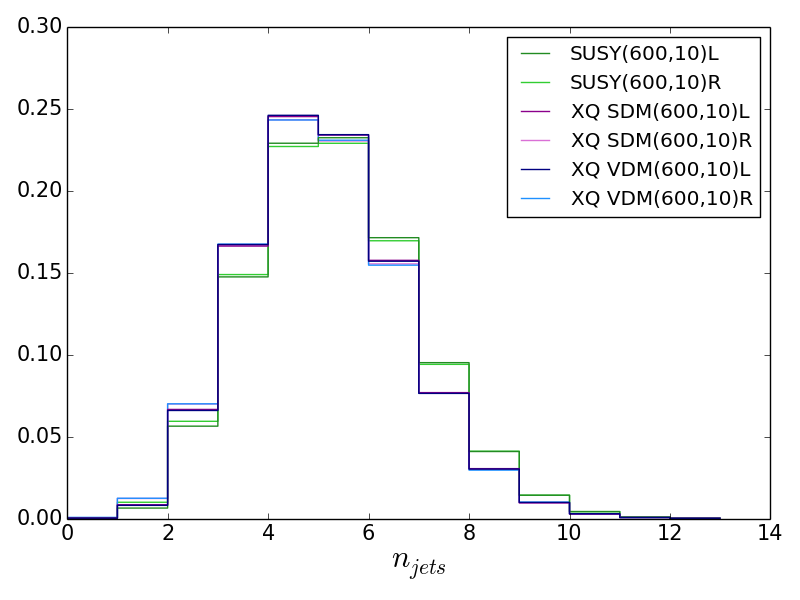}%
\includegraphics[width=0.34\textwidth]{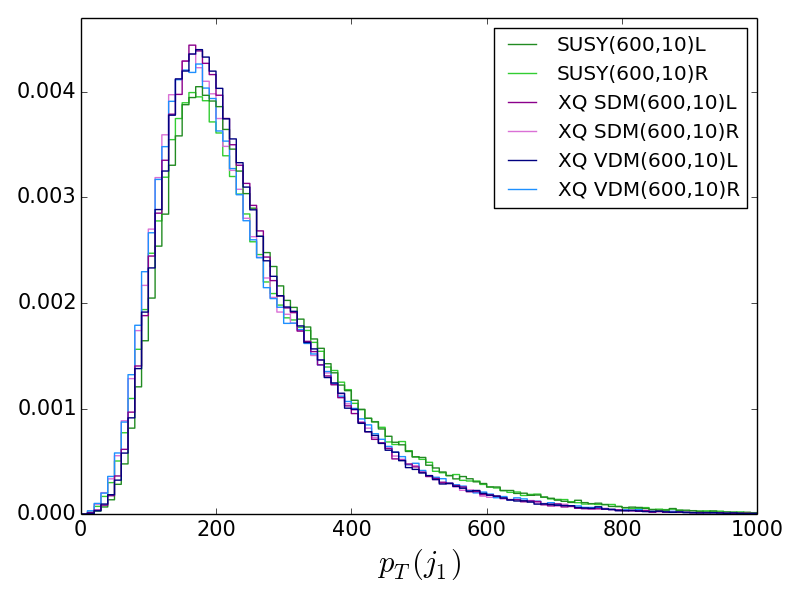}%
\includegraphics[width=0.34\textwidth]{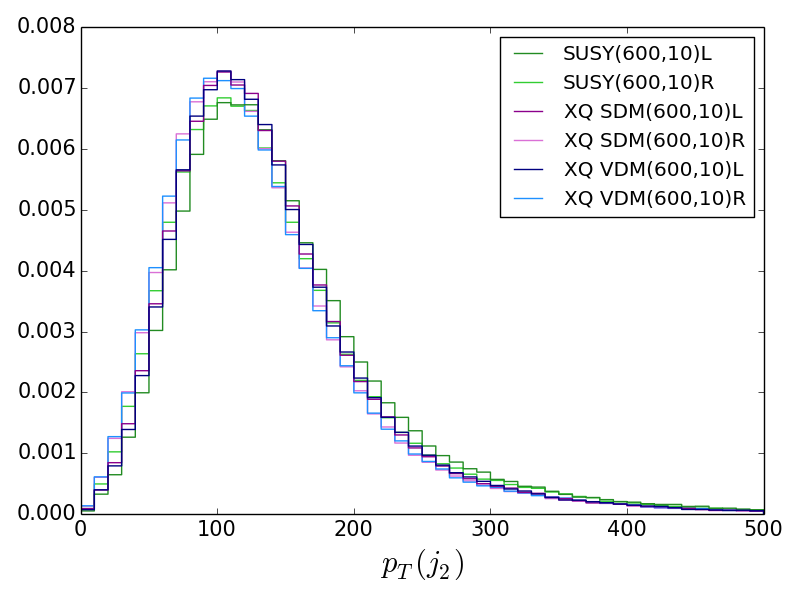}\\
\includegraphics[width=0.34\textwidth]{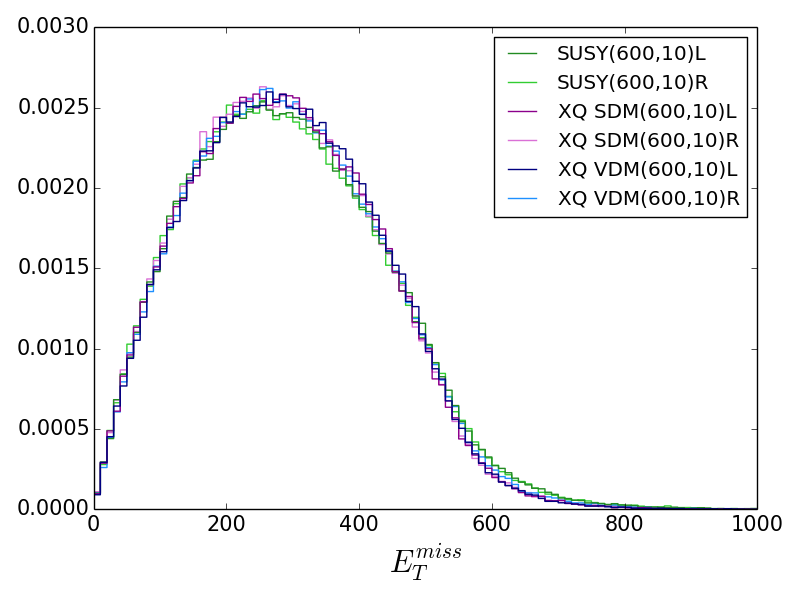}%
\includegraphics[width=0.34\textwidth]{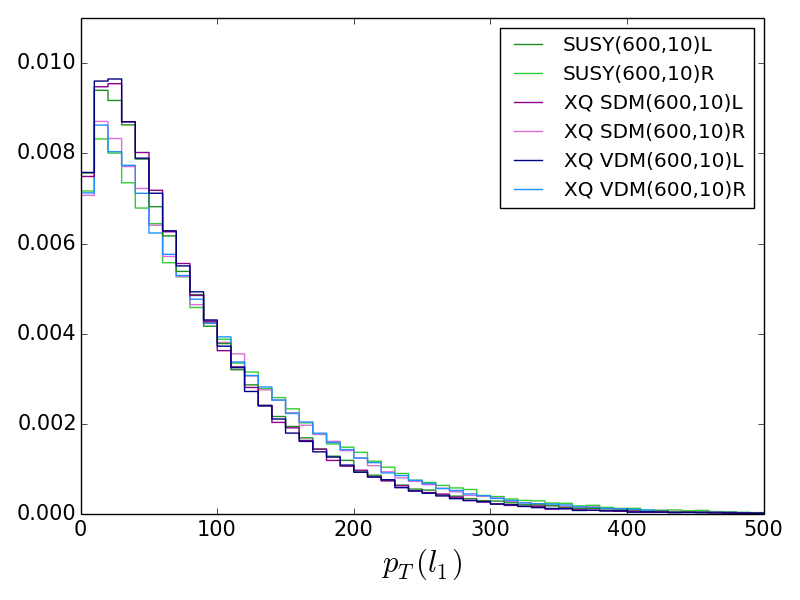}%
\includegraphics[width=0.34\textwidth]{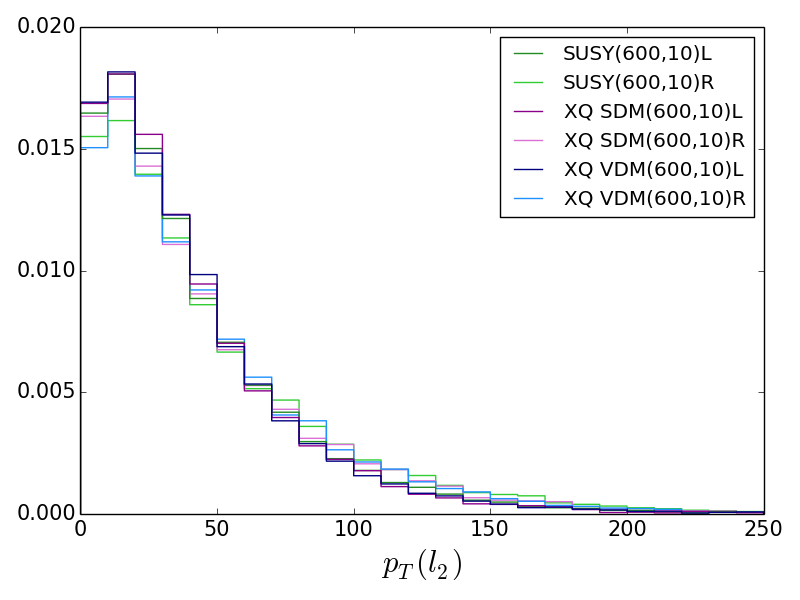}
\caption{ 
Differential distributions (normalized to one) of  jet multiplicity $n_{\rm jets}$, 
transverse momentum of the leading and sub-leading jet $p_T(j_1)$ and $p_T(j_2)$, 
missing transverse energy $\MET$, and $p_T$ of the leading and sub-leading lepton $p_T(l_1)$ and $p_T(l_2)$ 
for the mass combination $(600,\,10)$. 
}\label{fig:GenLevelDist1}
\end{figure}

As a first check whether we can expect specific differences in the cut efficiencies between the SUSY and XQ models, it is instructive to consider 
some basic parton-level distributions, as shown in Fig.~\ref{fig:GenLevelDist1} for the (600,\,10) mass combination. 
These distributions have been obtained using {\sc MadAnalysis}\,5 and considering the showered and hadronised event files from {\sc Pythia}; jets have been processed through {\sc FastJet}~\cite{Cacciari:2005hq,Cacciari:2011ma} using the anti-kt algorithm with minimum $p_T=5$~GeV and cone radius $R=0.5$. 
We see that the SUSY events tend to have more jets and a slightly harder $\MET$ spectrum. 
Moreover, the leading and sub-leading jets tend to be somewhat harder in the SUSY than in the XQ cases. 
Overall, these differences are however rather small and will likely not lead to any significant differences in the %SUSY and XQ 
cut efficiencies. 

\begin{figure}[h!]
\centering
\includegraphics[width=0.34\textwidth]{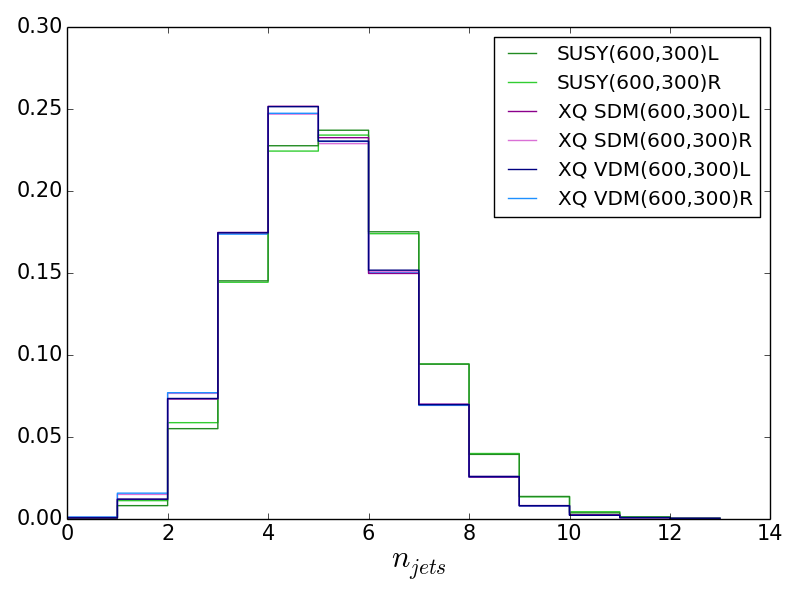}%
\includegraphics[width=0.34\textwidth]{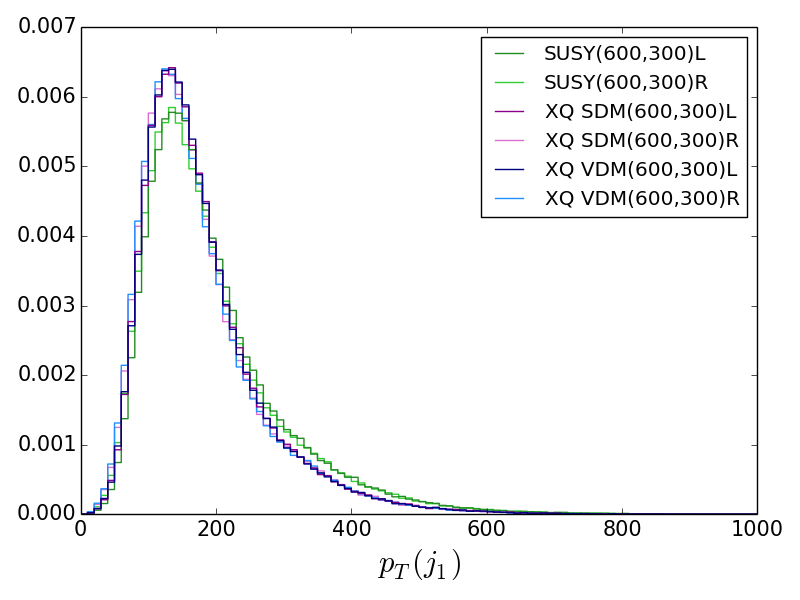}%
\includegraphics[width=0.34\textwidth]{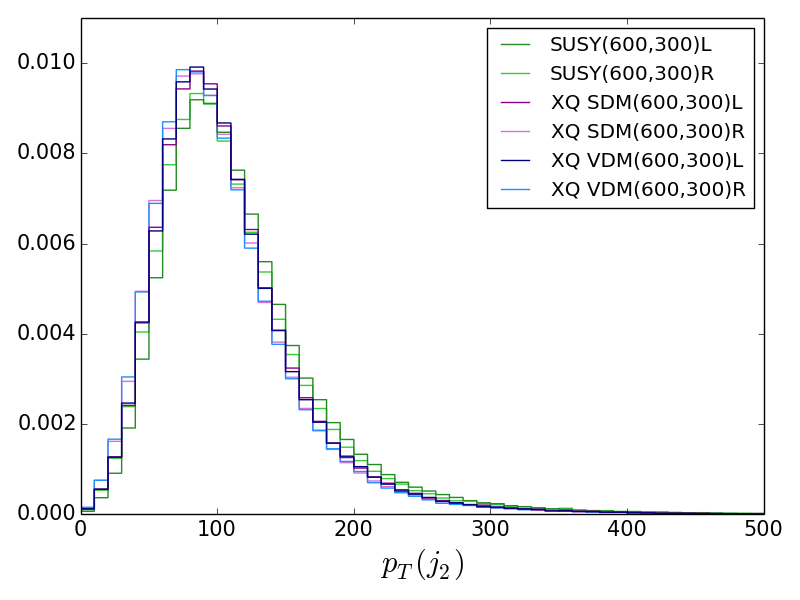}\\
\includegraphics[width=0.34\textwidth]{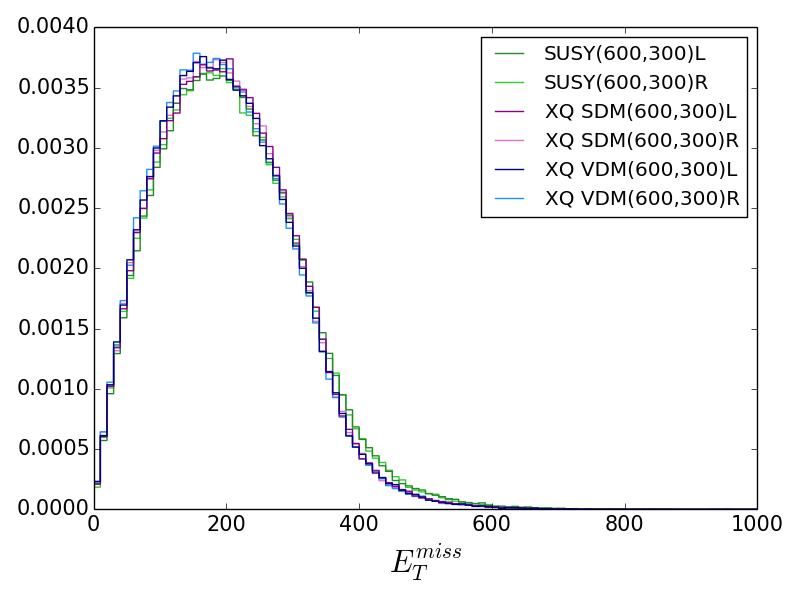}%
\includegraphics[width=0.34\textwidth]{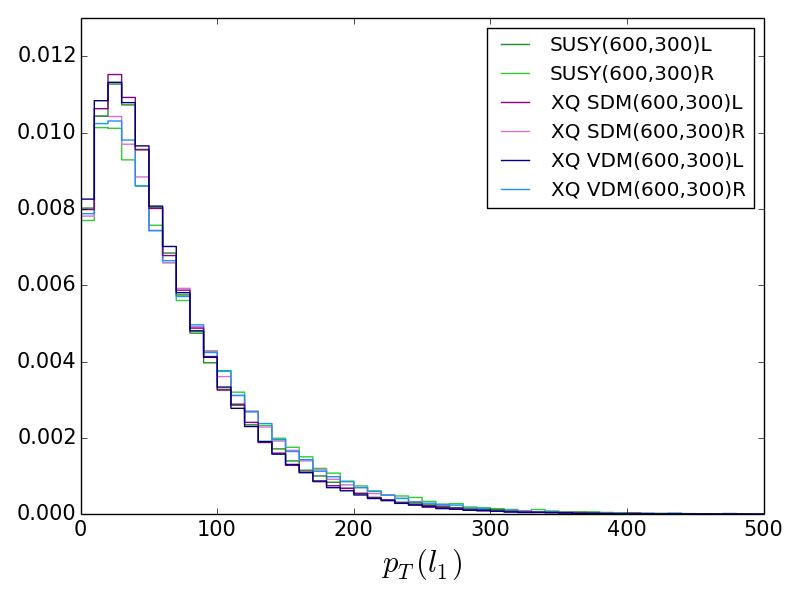}%
\includegraphics[width=0.34\textwidth]{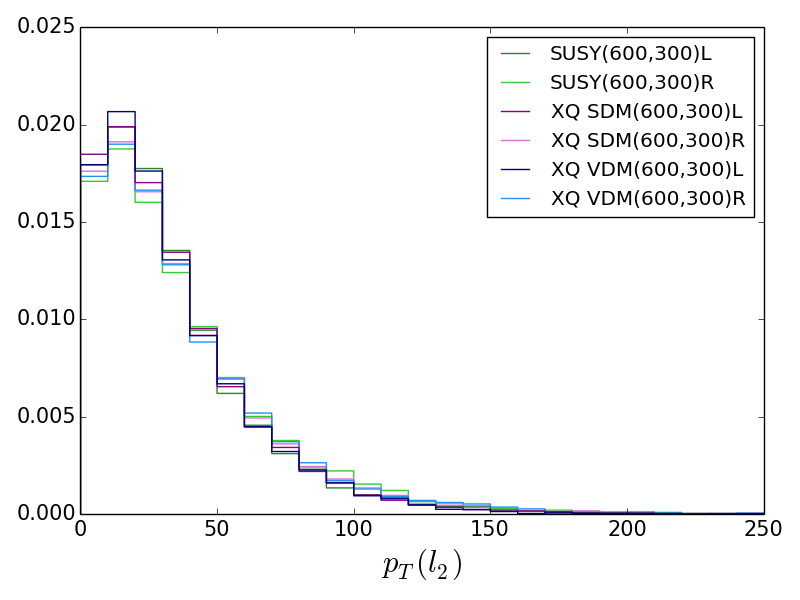}
\caption{ 
Same as Fig.~\ref{fig:GenLevelDist1} but for the $(600,\,300)$ mass combination. 
}\label{fig:GenLevelDist2}
\end{figure}

Regarding the lepton $p_T$, the small difference that appears is between the L and R cases 
rather than between SUSY and XQ: all the (600,\,10)R scenarios exhibit somewhat harder $p_T(l)$
than the (600,\,10)L scenarios. This comes from the fact that the top polarisation influences the $p_T$ of the top decay products.  %as discussed in \cite{Belanger:2012tm,Belanger:2013gha} and references therein. 
These features persist for smaller top-partner--DM mass difference, see Fig.~\ref{fig:GenLevelDist2}.

Polarisation effects in stop decays were studied in detail in \cite{Cao:2006wk,Nojiri:2008ir,Shelton:2008nq,Perelstein:2008zt,Berger:2012an,Chen:2012uw,Bhattacherjee:2012ir,Belanger:2012tm,Low:2013aza,Belanger:2013gha,Wang:2015ola}. Sizeable effects were found in kinematic distributions of the final-state leptons and $b$-quarks, and in particular in their angular correlations. While this might help to constrain the relevant mixing angles in precision studies of a positive signal~\cite{Perelstein:2008zt,Berger:2012an,Bhattacherjee:2012ir,Belanger:2012tm,Low:2013aza,Belanger:2013gha} and possibly to characterise the spin of the top-partner mediators and of the DM states through the structure of their coupling~\cite{Shelton:2008nq,Berger:2012an,Chen:2012uw}, as we will see, the current experimental analyses are not very sensitive to these effects. 

%\clearpage

\clearpage
%==============================================================================
\section{Effects in existing 8 TeV analyses}
\label{sec:analyses8tev}
%==============================================================================

Let us now analyse how the cut acceptances of existing 8 TeV analyses compare for the SUSY and XQ  scenarios. 
To this end, we consider the following ATLAS and CMS analyses implemented in {\sc CheckMATE}~\cite{Drees:2013wra} 
or the {\sc MadAnalysis}\,5 Public Analysis Database ({\sc MA5} PAD)~\cite{Dumont:2014tja}: 

\begin{itemize}
\item Fully hadronic stop search: ATLAS-CONF-2013-024~\cite{ATLAS:2013cma}  implemented in {\sc CheckMATE}, see Section~\ref{sec:fullhadstop}
\item Stop searches in the single lepton mode from ATLAS~\cite{Aad:2014kra}  ({\sc CheckMATE}) and CMS~\cite{Chatrchyan:2013xna} ({\sc MA5} PAD, recast code \cite{ma5code:cms-sus-13-011}), see Section~\ref{sec:1lepstop}
\item The stop search with 2 leptons from ATLAS~\cite{Aad:2014qaa} implemented in {\sc CheckMATE}, see Section~\ref{sec:2lepstop}
\item The generic gluino/squark search in the 2--6 jets plus missing energy channel from ATLAS~\cite{Aad:2014wea}  ({\sc MA5} PAD, recast code \cite{MA5-ATLAS-multijet-1405}), see Section~\ref{sec:genglusq}
\end{itemize}

%-----------------------------------------------------------------------------
\subsection{Fully hadronic stop search}
\label{sec:fullhadstop}
%-----------------------------------------------------------------------------

The ATLAS analysis~\cite{ATLAS:2013cma} implemented in {\sc CheckMATE} targets stop-pair production followed by stop decays into a top quark and the lightest neutralino, $pp\to \tilde t_1^{}\tilde t_1^*\to t \bar t \tilde\chi^0_1\tilde\chi^0_1$ 
in the fully-hadronic top final state, $t\to bW\to bq\bar q$. 
The search is thus conducted in events with large missing transverse momentum and six or more jets, of which $\ge 2$ must have been $b$-tagged. The two leading jets are required to have $p_T >80$~GeV with the remaining jets having $p_T >35$~GeV. Pre-selected electrons or muons, as well as taus are vetoed. Further requirements are imposed on azimuthal angle ($\Delta\phi$) and transverse mass ($m_T$) variables and on two 3-jet systems. 
Then three overlapping signal regions (SRs) are defined by requirements on $\ETmiss$, 
SR1:  $\ETmiss\ge 200$~GeV,  SR2:  $\ETmiss\ge 300$~GeV and SR3:  $\ETmiss\ge 350$~GeV.%
\footnote{We note that the conference note~\cite{ATLAS:2013cma} was superseded by the paper publication \cite{Aad:2014bva}, which has six SRs targeting the $\tilde t_1\to t\tilde\chi^0_1$ decay instead of three. Four of these, SRA1--4, are for ``fully resolved'' events with $\ge 6$ jets and a stacked $\ETmiss$ cut of 150, 250, 300 and 350~GeV. This is similar to the conference note. Two more SRs, SRB1--2, are for ``partially resolved'' events with 4 or 5 jets and higher $\ETmiss$, designed to target high stop masses.  Moreover, the paper considers three SRs, SRC1--3, optimized for stop decays into charginos. The limit is then set from a combination of SRA+B or SRA+C. Since this cannot be reproduced without a prescription of how to combine the SRs, 
we keep using the {\sc CheckMATE} implementation of the conference note to test the efficiencies of the hadronic stop search for our benchmark points. This is also justified by the fact that we are not primarily interested in the absolute limit but in potential differences in selection efficiencies between scalar and fermionic top partners.}

\begin{table}[t!]
\scriptsize
\centering %\hspace*{-8mm}
\begin{tabular}{lccc}
\hline
         & SUSY & XQ-SDM & XQ-VDM \\
\hline         
Initial no.\ of events & 200000 & 200000 & 200000 \\
$\ETmiss>80$~GeV (Trigger) & 187834 (-6.08 \%)  &  187872 (-6.06 \%)  & 188358 (-5.82 \%) \\ 
muon veto ($p_T>10$~GeV) & 154643 (-17.67 \%)  & 153946 (-18.06 \%) &  154710 (-17.86 \%) \\
electron veto ($p_T>10$~GeV) &  123420 (-20.19 \%)  & 122439 (-20.47 \%)  & 123247 (-20.34 \%) \\
$\ETmiss > 130$~GeV & 113638 (-7.93 \%)  & 112808 (-7.87 \%)  & 113620 (-7.81 \%) \\
$\ge 6$ jets, $p_T>80,80,35$~GeV & 33044 (-70.92 \%)  & 27987 (-75.19 \%)  & 28285 (-75.11 \%) \\
reconstr.\ $\ETmiss{^{\rm,track}} > 30$~GeV & 32564 (-1.45 \%)  & 27563 (-1.51 \%)  & 27901 (-1.36 \%) \\
$\Delta\phi$($\ETmiss, \ETmiss{^{\rm,track}}) < \pi/3$ & 31200 (-4.19 \%)  & 26583 (-3.56 \%)  & 26939 (-3.45 \%) \\
$\Delta\phi$($\ETmiss$, 3 hdst jets) > 0.2$\pi$ & 26276 (-15.78 \%)  &  22795 (-14.25 \%)  & 23129 (-14.14 \%) \\ 
tau veto & 22880 (-12.92 \%) & 19967 (-12.41 \%)  & 20354 (-12.00 \%) \\
2 $b$ jets & 9668 (-57.74 \%)  & 8510 (-57.38 \%) &  8660 (-57.45 \%) \\
$m_T(b\,{\rm jets}) > 175$~GeV & 7202 (-25.51 \%)  &  6447 (-24.24 \%)  & 6579 (-24.03 \%) \\
3 closest jets 80--270 GeV & 6437 (-10.62 \%)  & 5877 (-8.84 \%)  & 5929 (-9.88 \%) \\
same for second closest jets & 3272 (-49.17 \%) &  3186 (-45.79 \%)  & 3351 (-43.48 \%) \\
\hline
$\ETmiss\ge 150$~GeV & 3230 (-1.28 \%) &  3156 (-0.94 \%) &  3312 (-1.16 \%) \\
$\ETmiss\ge 200$~GeV (SR1) & 3067 (-5.05 \%) &  3000 (-4.94 \%) &  3161 (-4.56 \%) \\
$\ETmiss\ge 250$~GeV & 2795 (-8.87 \%) &  2732 (-8.93 \%) &  2867 (-9.30 \%) \\
$\ETmiss\ge 300$~GeV (SR2) & 2413 (-13.67 \%) & 2373 (-13.14 \%) & 2490 (-13.15 \%)\\
$\ETmiss\ge 350$~GeV (SR3) & 1948 (-19.27 \%) & 1926 (-18.84 \%) & 2010 (-19.28 \%)\\
\hline
\end{tabular}
\caption{\label{tab:cutflow-CM-1} Cut-flow of the hadronic stop analysis of ATLAS for Point~(600,\,10)L, derived with {\sc CheckMATE}.}
\end{table}

The effect of the various cuts is illustrated in Table~\ref{tab:cutflow-CM-1} for the example of Point (600,\,10)L. 
We observe that most preselection cuts have very similar efficiencies\footnote{Here and in the following, we use the term ``efficiency'' for the percentage of events remaining after one or more cuts. Strictly speaking this is the quantity acceptance$\times$efficiency, $A\epsilon$.}
when comparing SUSY and XQ cases. 
Small differences, of the level of few percent, occur only in the requirement of at least six jets (cf.\ Fig.~\ref{fig:GenLevelDist1}) and 
the condition on ``3 closest jets'' and ``second closest jets'', but these differences tend to compensate  each other. 
Finally, the effect of the $\ETmiss$ cuts that define the three SRs is almost the same for the SUSY and XQ scenarios. 
Consequently, the final numbers of events in each of the SRs agree within $\lesssim 5\%$ for the SUSY and XQ scenarios. 

The total efficiencies in the three SRs, cross-section excluded at 95\%~CL and corresponding top-partner mass limits in GeV are compared in Table~\ref{tab:hadr-eff-limits} for all four benchmark scenarios.\footnote{Given the upper limit on the cross-section together with the cross-section prediction as a function of the top-partner mass one can estimate the 95\% CL mass limit under the assumption that the efficiency is flat. While this kind of extrapolation is not a substitute for determining the true limit through a scan over the masses, it does give an indication of i) the impact of the differences in the excluded cross-section and ii) the higher reach in XQ as compared to SUSY. As we will see, this extrapolation works reasonably well for the stop searches 
but not for analyses that involve cuts which are directly sensitive to the overall mass scale.} 
We see that for a specific mass combination, the total efficiencies and hence the upper limit on the cross-section are very similar for the SUSY and XQ hypotheses.
The derived lower limit on the top-partner mass of course depends on the input cross-section (whether it is assumed SUSY-like or XQ-like), and is thus higher for the XQ interpretation than for the SUSY interpretation. 
However, the differences in the mass limits arising from applying SUSY, XQ-SDM or XQ-VDM efficiencies are generally small. Indeed for the (600,\,10) scenarios, i.e.\ large mass splitting, they are only 2--4~GeV, which is totally negligible. For smaller mass splittings, represented by the (600,\,300) scenarios, they reach about 10--20~GeV, which is still negligible.      
Finally, note that the effect on the mass limit from considering L vs.\ R polarised tops is of comparable size.

\begin{table}[t!]\centering
\scriptsize
\begin{tabular}{| l | ccc | ccc |}
\hline
& \multicolumn{3}{c|}{\bf Point~(600,\,10)L} &   \multicolumn{3}{c|}{\bf Point~(600,\,10)R}\\
\hline
         & SUSY & XQ-SDM & XQ-VDM & SUSY & XQ-SDM & XQ-VDM\\
\hline         
eff. SR1 & 0.015 & 0.015 & 0.016 & 0.014 & 0.015 & 0.014 \\
eff. SR2 & 0.012 & 0.012 & 0.012 & 0.011 & 0.012 & 0.011 \\
eff. SR3$^*$ & 0.0097 & 0.0096 & 0.010 & 0.0092 & 0.0095 & 0.0094 \\
\hline
excl. XS [pb] & 0.0196 & 0.0199 & 0.0189 & 0.0209 & 0.0201 & 0.0205 \\
mass limit/SUSY XS & 619 & 618 & 622 & 613 & 617 & 615 \\
mass limit/XQ XS & 805 & 803 & 808 & 798 & 802 & 800 \\
\hline
$1-{\rm CLs}$ & 0.98 & 1 & 1 & 0.97 & 1 & 1 \\
\hline
\end{tabular}\\[4mm]
\begin{tabular}{| l | ccc | ccc |}
\hline
& \multicolumn{3}{c|}{\bf Point~(600,\,300)L} &   \multicolumn{3}{c|}{\bf Point~(600,\,300)R} \\
\hline
         & SUSY & XQ-SDM & XQ-VDM & SUSY & XQ-SDM & XQ-VDM \\
\hline         
eff. SR1$^*$ & 0.0074 & 0.0064 & 0.0062  & 0.0066 & 0.0060 & 0.0053 \\
eff. SR2 & 0.0039 & 0.0032 & 0.0031  & 0.0035 & 0.0032 & 0.0026 \\
eff. SR3 & 0.0022 & 0.0016 & 0.0017  & 0.0018 & 0.0016 & 0.0013 \\
\hline
excl. XS [pb] & 0.0647 & 0.0759 & 0.0772 & 0.0726 & 0.0805 & 0.0910 \\
mass limit/SUSY XS & 522 & 510 & 509 & 514 & 506 & 497 \\
mass limit/XQ XS & 687 & 671 & 670 & 676 & 666 & 655 \\
\hline
$1-{\rm CLs}$ & 0.59  & 1 & 1 & 0.54 & 1 & 1 \\
\hline
\end{tabular}
\caption{\label{tab:hadr-eff-limits} Efficiencies in the three SRs, cross-section (XS) excluded at 95\%~CL, corresponding extrapolated top-partner mass limits in GeV, and CLs exclusion value from the hadronic stop analysis of ATLAS derived with {\sc CheckMATE}. ``mass limit/SUSY XS'' means that the excluded XS is translated to a mass limit using the SUSY production cross-section from Fig.~\ref{fig:XS}, while ``mass limit/XQ XS'' means the limit is estimated using the XQ cross-section. The exclusion CL is obtained considering the corresponding cross-sections at 600 GeV, $\sigma(\st_1\st_1^*)=0.024$~pb for stop production and $\sigma(T\bar T)=0.167$~pb for XQ production. The most sensitive SR used for the limit setting is marked with a star.}
\end{table}

%\clearpage
%-----------------------------------------------------------------------------
\subsection{Stop search in the single lepton final state}
\label{sec:1lepstop}
%-----------------------------------------------------------------------------

Stops are also searched for in final states with a single lepton, jets and $\ETmiss$, arising from one $W$ decaying leptonically while the other one decays hadronically. The ATLAS analysis \cite{Aad:2014kra} for this channel is implemented in {\sc CheckMATE}, while the (cut-based version of) the corresponding CMS analysis~\cite{Chatrchyan:2013xna} is implemented in the {\sc MA5} PAD. 

In the CMS analysis~\cite{Chatrchyan:2013xna}, events are required to contain one isolated electron (muon) 
with $p_T > 30$ (25) GeV, no additional isolated track or hadronic $\tau$ candidate, at least four jets with 
$p_T > 30$~GeV at least one of which must be $b$-tagged, $\ETmiss > 100$~GeV and $M_T>120$~GeV. 
The analysis further makes use of the quantity $M_{\rm T2}^W$, a hadronic top $\chi^2$ ensuring that  three of the jets in the event be consistent with the $t\to bW\to bq\bar q$ decay, and the topological variable $\Delta\phi(\ETmiss,\, {\rm jet})$. Various signal regions are defined targeting $\tilde t_1^{}\to t \tilde\chi^0_1$ or $\tilde t_1^{}\to b \tilde\chi^+_1$ decays with small or large mass differences between the stop and the neutralino or chargino. 

As an illustrative example, we show in Table~\ref{tab:cutflow-MA5-1lep} the cut-flow for the ``$\tilde t_1^{}\to t \tilde\chi^0_1$, high $\Delta M$, $\ETmiss>300$~GeV'' signal region for Point~(600,\,10)R, which is the most sensitive SR for this benchmark.  
The only noticeable difference, though hardly of the level of 5\% in the cut efficiency, arises from the requirement of at least four jets. All other cuts have again almost the same effects on the SUSY and XQ models. 
Altogether, starting from the same number of events, we end up with slightly more SUSY than XQ events in this SR, but this difference is only 6--7\%. 

\begin{table}[t!]
\scriptsize
\centering 
\begin{tabular}{lccc}
\hline
         & SUSY & XQ-SDM & XQ-VDM \\
\hline         
Initial no.\ of events & 200000 & 200000 & 200000 \\
$\ge1$ candidate lepton & 51097 (-74.45 \%) & 50700 (-74.65 \%) & 50417 (-74.79 \%) \\
$\ge4$ central jets & 23737 (-53.55 \%) & 21333 (-57.92 \%) & 20997 (-58.35 \%)  \\
$\ETmiss > 50$~GeV & 23203 (-2.25 \%) & 20848 (-2.27 \%) & 20548 (-2.14 \%)  \\
$\ETmiss > 100$~GeV & 21640 (-6.74 \%) & 19393 (-6.98 \%) & 19206 (-6.53 \%)  \\
$\ge 1~b$-tagged jet & 18339 (-15.25 \%) & 16643 (-14.18 \%) & 16512 (-14.03 \%)  \\
isol lepton and track veto & 17370 (-5.28 \%) & 15892 (-4.51 \%) & 15750 (-4.61 \%)  \\
hadronic tau veto & 17061 (-1.78 \%) & 15646 (-1.55 \%) & 15487 (-1.67 \%)  \\
$M_T > 120$~GeV & 13811 (-19.05 \%) & 12788 (-18.27 \%) & 12691 (-18.05 \%)  \\
$\Delta\phi$($\ETmiss$, j1 or j2) > 0.8 & 12006 (-13.07 \%) & 11251 (-12.02 \%) & 11164 (-12.03 \%)  \\
$\chi^2 < 5$ & 7079 (-41.04 \%) & 6771 (-39.82 \%) & 6750 (-39.54 \%)  \\
$\ETmiss > 300$~GeV & 4138 (-41.55 \%) & 3820 (-43.58 \%) & 3929 (-41.79 \%)  \\
$M_{\rm T2}^W > 200$~GeV & 3030 (-26.78 \%) & 2830 (-25.92 \%) & 2851 (-27.44 \%)  \\
\hline
\end{tabular}
\caption{\label{tab:cutflow-MA5-1lep} Cut-flow for the ``$\tilde t_1^{}\to t \tilde\chi^0_1$, high $\Delta M$, $\ETmiss>300$~GeV'' signal region (denoted SR-A) of the CMS stop search in the 1-lepton channel for Point~(600,\,10)R, derived with the {\sc MadAnalysis}\,5 recast code~\cite{ma5code:cms-sus-13-011}. 
Note that the event weighting to account for trigger and lepton identification efficiencies and for initial-state radiation effects is not included in this cut-flow. More details about these aspects and their implementation of the recast code can be found in the original references~\cite{Chatrchyan:2013xna} and \cite{ma5code:cms-sus-13-011}.} 
\end{table}

\begin{table}[t!]\centering
\scriptsize
\begin{tabular}{| l | ccc | ccc |}
\hline
& \multicolumn{3}{c|}{\bf Point~(600,\,10)L} &   \multicolumn{3}{c|}{\bf Point~(600,\,10)R} \\
\hline
         & SUSY & XQ-SDM & XQ-VDM & SUSY & XQ-SDM & XQ-VDM \\
\hline         
eff. SR-A & 0.0108 & 0.0109 & 0.0111 & 0.0108$^*$ & 0.0106$^*$ & 0.0107$^*$ \\
eff. SR-B & 0.0181$^*$ & 0.0176$^*$ & 0.0184$^*$ & 0.0154 & 0.0152 & 0.0153 \\
\hline
excl. XS [pb] & 0.0169 & 0.0173 & 0.0166 & 0.0210 & 0.0213 & 0.0211 \\
mass limit/SUSY XS & 631 & 629 & 633 & 613 & 611 & 612 \\
mass limit/XQ XS & 820 & 818 & 822 & 798 & 796 & 797 \\
\hline
$1-{\rm CLs}$ &0.99 & 1 & 1 & 0.97 & 1 & 1 \\
\hline
\end{tabular}\\[4mm]
\begin{tabular}{| l | ccc | ccc |}
\hline
& \multicolumn{3}{c|}{\bf Point~(600,\,300)L} &   \multicolumn{3}{c|}{\bf Point~(600,\,300)R} \\
\hline
         & SUSY & XQ-SDM & XQ-VDM & SUSY & XQ-SDM & XQ-VDM \\
\hline         
eff. SR-A & \!0.00360\! & 0.00366 & 0.00346  & \!0.00340\! & 0.00321 & 0.00315 \\
eff. SR-B& \!0.00748\!$^*$ & 0.00685$^*$ & 0.00632$^*$  & \!0.00597\!$^*$ & 0.00570$^*$ & 0.00536$^*$ \\
\hline
excl. XS [pb] & 0.0399 & 0.0448 & 0.0480 & 0.0507 & 0.0530 & 0.0563 \\
mass limit/SUSY XS & 560 & 551 & 546 & 541 & 538 & 533 \\
mass limit/XQ XS & 733 & 722 & 715 & 710 & 706 & 700 \\
\hline
$1-{\rm CLs}$ &0.81 & 1 & 1 & 0.72 & 1 & 1 \\
\hline
\end{tabular}
\caption{\label{tab:lept-eff-limits-cms} Efficiencies for the 
``$\tilde t_1^{}\to t \tilde\chi^0_1$, high $\Delta M$, $\ETmiss>300$~GeV'' (denoted SR-A) and 
``$\tilde t_1^{}\to b \tilde\chi^+_1$, high $\Delta M$, $\ETmiss>250$~GeV'' (denoted SR-B) signal regions, 
cross-sections excluded at 95\%~CL, corresponding extrapolated top-partner mass limits in GeV, and CLs exclusion value 
from the 1-lepton stop analysis of CMS, derived with the {\sc MadAnalysis}\,5 recast code~\cite{ma5code:cms-sus-13-011}. The most sensitive SR used for the limit setting is indicated by a star. }
\end{table}

Table~\ref{tab:lept-eff-limits-cms} summarises the total efficiencies in the two most important SRs of this analysis,
the cross-sections excluded at 95\%~CL and the corresponding top-partner mass limits in GeV for all four benchmark scenarios.
Note that, for large mass splitting, the SRs ``$\tilde t_1^{}\to b \tilde\chi^+_1$, high $\Delta M$, $\ETmiss>250$~GeV'' (here denoted as SR-B) which is optimized for $\tilde t_1^{}\to b \tilde\chi^+_1$ decays and ``$\tilde t_1^{}\to t \tilde\chi^0_1$, high $\Delta M$, $\ETmiss>300$~GeV'' (denoted SR-A) optimized for $\tilde t_1^{}\to t \tilde\chi^0_1$ have very similar sensitivities. 
In fact we observe that the most sensitive SR depends on the top polarisation. 
Events with left polarised tops are more likely to pass the additional requirement of SR-B on the leading $b$-jet, $p_T > 100$~GeV.
Concretely, in the SUSY scenario the expected upper limits are $0.0290$ pb in SR-A versus $0.0251$ pb in SR-B for (600,10)L and $0.0291$ pb vs.\ $0.0295$ pb for (600,10)R.
CMS has observed a small underfluctuation in both these SRs: 2 observed events vs.
 $4.7 \pm 1.4$ expected in SR-A and 5 observed events vs.\ $9.9 \pm 2.7$ expected in SR-B.
Overall the observed cross-section limit is somewhat lower in the left-polarised scenario.
An analogous observation holds for the XQ scenarios; the differences between SUSY and XQ scenarios are negligible.

Finally, for smaller mass gaps, SR-B is more sensitive in all considered scenarios and we observe differences at the level of 10--15\% in the total signal selection efficiencies, which translate into up to about 20\% differences in the excluded cross-sections, 
or $\lesssim 5\%$ in the estimated mass limits. 
The uncertainty from considering scenarios that lead to left or right polarised tops is of similar magnitude.
The latter is consistent with the observation in \cite{Chatrchyan:2013xna} that the limits on the $\tilde t_1^{}$ and $\tilde\chi^0_1$ masses vary by $\pm10$--$20$~GeV depending on the top-quark polarisation; the polarisation dependence in the $\tilde t_1^{}\to b\tilde\chi^+_1$ channel can be somewhat larger. 

\bigskip

The corresponding ATLAS search \cite{Aad:2014kra} for this channel is implemented in {\sc CheckMATE}. 
Here, the signal selection requires a least one ``baseline'' lepton with $p_T>10$~GeV, which is later tightened 
to exactly one isolated lepton with $p_T > 25$~GeV.\footnote{Except for the SR with soft-lepton
selections which employ a $p_T$ > 6(7) GeV requirement for muons (electrons).} Events containing additional
baseline leptons are rejected.  
The analysis comprises 15 non-exclusive SRs, 4 of which target $\tilde t_1^{}\to t \tilde\chi^0_1$ (labelled `{\tt tN\_}'), 9 target $\tilde t_1^{}\to b \tilde\chi^+_1$ (labelled `{\tt bC\_}'), and the last 2 target 3-body and mixed decays. 
A minimum number of jets ranging between 2 and 4 is required depending on the SR, together with $b$-tagging requirements and an $\ETmiss$ cut of at least 100~GeV. 
As for the CMS analysis, a number of kinematic variables 
($m_T$, $am_{T2}$, $\Delta\phi(\ETmiss,\vec p_T({\rm jet}))$, etc.) are exploited for reducing the background. 
The relevant SRs for our benchmark points are {\tt tN\_med}, {\tt bCd\_high} and {\tt bCd\_bulk}.\footnote{Note that the ATLAS search has a dedicated SR to target boosted final states, {\tt tN\_boost}. This SR is not considered here, as the relevant ``topness'' variable is not implemented in {\sc CheckMATE}.} 
Of course, for the limit setting only the most sensitive one is used.  
A partial cut-flow example is given in Table~\ref{tab:cutflow-CM-1lep} for Point (600,\,10)R. 
The results for all four benchmark points are summarised in Table~\ref{tab:lept-eff-limits-atlas-corrected}.

\begin{table}[t!]\centering 
\scriptsize
\begin{tabular}{lccc}
\hline
         & SUSY & XQ-SDM & XQ-VDM \\
\hline        
Initial no.\ of events & 200000 & 200000 & 200000 \\ 
Trigger & 158881 (-20.56 \%) & 158929 (-20.54 \%) & 160073 (-19.96 \%)  \\
DQ & 154759 (-2.59 \%) & 155073 (-2.43 \%) & 156148 (-2.45 \%)  \\
\hline
1 baseline electron &  30142 (-80.52 \%) & 29980 (-80.67 \%) & 30019 (-80.78 \%) \\
1 signal electron &  22342 (-25.88 \%) & 22177 (-26.03 \%) & 22169 (-26.15 \%) \\
$\ge3$ jets $p_T\ge25$~GeV &  19865 (-11.09 \%) & 19241 (-13.24 \%) & 19262 (-13.11 \%) \\
$\ge4$ jets $p_T\ge25$~GeV &  14458 (-27.22 \%) & 13275 (-31.01 \%) & 13355 (-30.67 \%)  \\
\ldots & & & \\
{\tt tN\_med} $e$ & 1892 (-86.91 \%) & 1951 (-85.30 \%) & 1987 (-85.12 \%) \\
{\tt bCd\_high1} $e$ & 1792 (-87.61 \%) & 1651 (-87.56 \%) & 1748 (-86.91 \%) \\
{\tt bCd\_bulk} $e$ & 4359 (-69.85 \%) & 4180 (-68.51 \%) & 4262 (-68.09 \%) \\
\hline 
1 baseline $\mu$  & 27993 (-81.91 \%) & 28381 (-81.70 \%) & 28119 (-81.99 \%)  \\
1 signal $\mu$ &  23123 (-17.40 \%) & 23383 (-17.61 \%) & 23088 (-17.89 \%) \\ 
$\ge3$ jets $p_T\ge25$~GeV &  20695 (-10.50 \%) & 20624 (-11.80 \%) & 20302 (-12.07 \%)  \\
$\ge4$ jets $p_T\ge25$~GeV &  15197 (-26.57 \%) & 14448 (-29.95 \%) & 14163 (-30.24 \%) \\
\ldots & & & \\
{\tt tN\_med} $\mu$ & 2108 (-86.13 \%) & 1970 (-86.36 \%) & 1977 (-86.04 \%) \\
{\tt bCd\_high1} $\mu$ & 1790 (-88.22 \%) & 1821 (-87.40 \%) & 1747 (-87.67 \%) \\
{\tt bCd\_bulk} $\mu$ & 4582 (-69.85 \%) & 4415 (-69.44 \%) & 4340 (-69.36 \%) \\
\hline
\end{tabular}
\caption{\label{tab:cutflow-CM-1lep} Partial cut-flows for the ATLAS stop search in the 1-lepton channel for Point~(600,\,10)R, derived with {\sc CheckMATE}. Shown are the effects of the preselection cuts and the final numbers of events in specific signal regions. The cut-flows are given separately for electrons and muons.}
\end{table}

\begin{table}[t!]\centering
\scriptsize
\begin{tabular}{| l | ccc | ccc |}
\hline
& \multicolumn{3}{c|}{\bf Point~(600,\,10)L} &   \multicolumn{3}{c|}{\bf Point~(600,\,10)R} \\
\hline
         & SUSY & XQ-SDM & XQ-VDM & SUSY & XQ-SDM & XQ-VDM \\
\hline
eff. bCd{\tiny \_}bulk{\tiny \_}d & 0.0298* & 0.0287  & 0.0297  & 0.0278* & 0.0264* & 0.0270*\\
eff. bCd{\tiny \_}high1 & 0.0208  & 0.0204* & 0.0210* & 0.0179  & 0.0174  & 0.0175 \\
\hline
excl. XS [pb] & 0.0250 &  0.0335 & 0.0324 & 0.0267 & 0.0281 & 0.0274\\
mass limit/SUSY XS & 598 &  574 &  577 & 593 & 588 & 590 \\
mass limit/XQ XS & 780 & 750 & 754 & 773 & 768 & 770 \\
\hline
$1-{\rm CLs}$ &0.94 & 1 & 1 & 0.93 & 1 & 1 \\
\hline
\end{tabular}\\[4mm]
\begin{tabular}{| l | ccc | ccc |}
\hline
& \multicolumn{3}{c|}{\bf Point~(600,\,300)L} &   \multicolumn{3}{c|}{\bf Point~(600,\,300)R} \\
\hline
         & SUSY & XQ-SDM & XQ-VDM & SUSY & XQ-SDM & XQ-VDM \\
\hline
eff. bCd{\tiny \_}high1 & 0.00919* & 0.00810* & 0.00761* & 0.00777  & 0.00691  & 0.00638 \\
eff. tN{\tiny \_}med & 0.00927     & 0.00869  & 0.00836  & 0.00877* & 0.00862* & 0.00775*\\
\hline
excl. XS [pb] & 0.0742 & 0.0845& 0.0898 & 0.0509 & 0.0517 & 0.0579\\
mass limit/SUSY XS & 512 & 502 &  498 & 541 & 540 & 531 \\
mass limit/XQ XS & 673 & 661 & 656 & 709 & 708& 697\\
\hline
$1-{\rm CLs}$ &0.35 & 1 & 1 & 0.69 & 1 & 1 \\
\hline
\end{tabular}
\caption{\label{tab:lept-eff-limits-atlas-corrected}  Efficiencies for selected SRs, cross-sections excluded at 95\%~CL , corresponding extrapolated top-partner mass limits in GeV, and CLs exclusion values for the ATLAS stop search in the 1-lepton channel, derived with {\sc CheckMATE}. The most sensitive SR used for the limit setting is indicated by a star.}
\end{table}

As in the CMS analysis, we observe very similar sensitivities in several signal regions, 
and it depends on details of the scenario which SR turns out as the best one. 
It should be noted here that small differences in selection efficiencies can have a considerable impact on the observed limit if they yield different SRs as the most sensitive one. 
In particular, ATLAS has observed more events than expected in SR {\tt bCd\_high1} (16 observed events vs.\ $11 \pm 1.5$ expected). Consequently, limits obtained from this SR are weaker than those using {\tt tN\_med} (12 observed vs.\ $13 \pm 2.2$ expected) or {\tt bCd\_bulk\_d} (29 observed vs.\ $26.5 \pm 2.6$ expected). 
This is relevant, for example, for Point (600,\,10)L. 
Nonetheless, the differences when comparing SUSY, XQ-SDM and XQ-VDM cases remain small, 
in particular always well below the 20--30\% estimated systematic uncertainty inherent to recasting with fast simulation tools.
It is also worth pointing out that, in contrast to its CMS counterpart, this ATLAS analysis tends to give stronger limits for  R than for L scenarios. The effect is more pronounced for smaller mass differences, in agreement with Fig.~24 in \cite{Aad:2014kra}. Overall, the sensitivity to polarisation effects, while larger than for the CMS analysis, remains small.

%\clearpage
%-----------------------------------------------------------------------------
\subsection{Stop search in the 2-leptons final state}
\label{sec:2lepstop}
%-----------------------------------------------------------------------------

Let us next discuss the 2-lepton final state considered in the ATLAS analysis~\cite{Aad:2014qaa}. 
This analysis searches for direct stop-pair production with $\tilde t_1^{}\to b \tilde\chi^+_1\to bW^{(*)}\tilde\chi^0_1$ or 
$\tilde t_1^{}\to t \tilde\chi^0_1\to bW\tilde\chi^0_1$, targeting leptonic $W$ decays. 
Events are required to have exactly two oppositely charged signal leptons
(electrons, muons or one of each, defining same flavour (SF) and different-flavour (DF) selections). 
At least one of these electrons or muons must have
$p_T > 25$~GeV and $m_{\ell\ell} > 20$~GeV.
Events with a third preselected electron or muon are rejected. 
The analysis is subdivided into a ``leptonic mT2'' and ``hadronic mT2''  analysis, as well a 
multivariate analysis (MVA), which cannot be reproduced with our simulation frameworks. 
The ``leptonic mT2'' (4 SRs) and ``hadronic mT2'' (1 SR) analyses respectively use $m_{T2}$ and $m_{T2}^{b{\rm -jet}}$ 
as the key discriminating variable. Other kinematic variables used include
$\Delta\phi_j$ ($\Delta\phi_\ell$), the azimuthal angular distance between the $p_T^{\rm miss}$
vector and the direction of the closest jet (highest $p_T$ lepton).

The ``leptonic mT2'' analysis has 4 overlapping SRs defined by $m_{T2}>90$, 100, 110 and 120~GeV. 
From these, seven statistically independent SRs denoted S1--S7 are defined in the (jet selections, $m_{T2})$ plane, 
where `jet selections' refers to the number of jets with a certain minimum $p_T$, see Fig.~13 in \cite{Aad:2014qaa}. 
The most sensitive one for our benchmark points is S5, which has $m_{T2}>120$~GeV and at least two jets with 
$p_T({\rm jet1})>100$~GeV and $p_T({\rm jet2})>50$~GeV. 

Table~\ref{tab:cutflow-CM-2lep} shows a cut-flow example for the SF selection for Point~(600,\,10)R, 
as well as an abbreviated version for the DF selection.
Note that the leptonic $W$ decay was enforced in {\sc Pythia} to increase statistics. 
The SF selection gives less events than the DF one because the $Z$ veto removes about 20\% of events in the former but none in the latter. 
The combined count for SR S5 is given as the last line in the table. 
As was already the case for the other analyses, no significant differences occur at any particular step of the cut-flow. 
At the end we are left with the marginal difference of 4\% more XQ than SUSY events in a total selection efficiency 
of barely 3 permil (when considering events where the W is allowed to decay to anything).
% after folding the $W$ decay branching ratios back in). 

The picture is similar for Point~(600,\,10)L, for which %an abbreviated 
the cut-flow is given in Table~\ref{tab:cutflow-CM-2lep-Left}. 
Noteworthy is the fact that the initial difference in Points (600,\,10)R and (600,\,10)L from the 2 lepton selection (the first cut) is inverted by the last cut, so that in the final SR there remain more events for (600,\,10)L than for (600,\,10)R. 
This is a consequence of the dependence on the top polarisation already noted in the parton-level plots in Figs.~\ref{fig:GenLevelDist1} and \ref{fig:GenLevelDist2}. 

Either way, as can be seen from Table~\ref{tab:lept-eff-limits-atlas}, there is again no significant difference %whatsoever
in the total efficiencies and excluded cross-sections between SUSY, XQ-SDM and XQ-VDM scenarios. 

\begin{table}[t!]\centering 
\scriptsize
\begin{tabular}{lccc}
\hline
         & SUSY & XQ-SDM & XQ-VDM \\
\hline        
Initial no.\ of events & 200000 & 200000 & 200000 \\ 
2 leptons, $p_T>10$~GeV &  63129 (-68.44 \%) & 63877 (-68.06 \%) & 63604 (-68.20 \%) \\
\hline\hline
same flavour &  31464 (-50.16 \%) & 32040 (-49.84 \%) & 31643 (-50.25 \%) \\
isolation & 28096 (-10.70 \%) & 28538 (-10.93 \%) & 28234 (-10.77 \%) \\
opposite sign &  27961 (-0.48 \%) & 28402 (-0.48 \%) & 28078 (-0.55 \%) \\
$m_{\ell\ell} > 20$~GeV & 27457 (-1.80 \%) & 27874 (-1.86 \%) & 27586 (-1.75 \%) \\
$p_T(\ell) > 25$~GeV & 26505 (-3.47 \%) & 26948 (-3.32 \%) & 26625 (-3.48 \%) \\
$Z$ veto &  21448 (-19.08 \%) & 21682 (-19.54 \%) & 21374 (-19.72 \%) \\
$\Delta\phi_j>1$ & 12664 (-40.95 \%) & 13463 (-37.91 \%) & 13375 (-37.42 \%) \\
$\Delta\phi_b<1.5$ &  11779 (-6.99 \%) & 12638 (-6.13 \%) & 12460 (-6.84 \%) \\
$m_{T2}>120$~GeV &  4824 (-59.05 \%) & 5441 (-56.95 \%) & 5368 (-56.92 \%) \\
\hline
S5 -- SF (2 jets, $p_T>100,50$~GeV) & 2378 (-50.70 \%) & 2621 (-51.83 \%) & 2446 (-54.43 \%)\\
\hline\hline
different flavour & 31665 (-49.84 \%) & 31837 (-50.16 \%) & 31961 (-49.75 \%) \\
... & & & \\
$m_{T2}>120$~GeV &  5955 (-59.74 \%) & 6515 (-58.31 \%) & 6697 (-57.45 \%) \\
S5 -- DF (2 jets, $p_T>100,50$~GeV) & 3032 (-49.08 \%) & 3013 (-53.75 \%) & 3030 (-54.76 \%)\\
\hline\hline
S5 -- SF+DF & 5410 & 5634 & 5476\\
\hline
\end{tabular}
\caption{\label{tab:cutflow-CM-2lep} Cut-flow example for the ATLAS stop search in the 2-lepton channel for Point~(600,\,10)R, derived with {\sc CheckMATE}. Here, the leptonic $W$ decay was enforced to enhance statistics.}
\end{table}

\begin{table}[t!]\centering 
\scriptsize
\begin{tabular}{lccc}
\hline
         & SUSY & XQ-SDM & XQ-VDM \\
\hline        
Initial no.\ of events & 200000 & 200000 & 200000 \\ 
2 leptons, $p_T>10$~GeV &  60379 (-69.81 \%) & 61193 (-69.40 \%) & 60812 (-69.59 \%)  \\
\hline\hline
same flavour &   30109 (-50.13 \%) & 30508 (-50.14 \%) & 30419 (-49.98 \%) \\
isolation & 26759 (-11.13 \%) & 27108 (-11.14 \%) & 27066 (-11.02 \%) \\
opposite sign &  26660 (-0.37 \%) & 26994 (-0.42 \%) & 26987 (-0.29 \%) \\
$m_{\ell\ell} > 20$~GeV & 26043 (-2.31 \%) & 26364 (-2.33 \%) & 26381 (-2.25 \%) \\
$p_T(\ell) > 25$~GeV & 25062 (-3.77 \%) & 25251 (-4.22 \%) & 25345 (-3.93 \%) \\
$Z$ veto &  19570 (-21.91 \%) & 19765 (-21.73 \%) & 19642 (-22.50 \%) \\
$\Delta\phi_j>1$ & 11797 (-39.72 \%) & 12485 (-36.83 \%) & 12522 (-36.25 \%) \\
$\Delta\phi_b<1.5$ &  11270 (-4.47 \%) & 11943 (-4.34 \%) & 12035 (-3.89 \%) \\
$m_{T2}>120$~GeV &  4390 (-61.05 \%) & 4785 (-59.93 \%) & 4815 (-59.99 \%)  \\
\hline
S5 -- SF (2 jets, $p_T>100,50$~GeV) & 2711 (-38.25 \%) & 2803 (-41.42 \%) & 2841 (-41.00 \%) \\
\hline\hline
different flavour &  30270 (-49.87 \%) & 30685 (-49.86 \%) & 30393 (-50.02 \%) \\
... & & & \\
$\Delta\phi_j>1$ & 15273 (-38.59 \%) & 16117 (-36.31 \%) & 15896 (-36.21 \%)  \\
$\Delta\phi_b<1.5$ &  14683 (-3.86 \%) & 15505 (-3.80 \%) & 15260 (-4.00 \%) \\
$m_{T2}>120$~GeV &  5581 (-61.99 \%) & 6149 (-60.34 \%) & 5985 (-60.78 \%) \\
S5 -- DF (2 jets, $p_T>100,50$~GeV) & 3524 (-36.86 \%) & 3562 (-42.07 \%) & 3503 (-41.47 \%) \\
\hline\hline
S5 -- SF+DF & 6235 & 6365 & 6344\\
\hline
\end{tabular}
\caption{\label{tab:cutflow-CM-2lep-Left} Cut-flow example for the ATLAS stop search in the 2-lepton channel for Point~(600,\,10)L, derived with {\sc CheckMATE}. To be compared with Table~\ref{tab:cutflow-CM-2lep}.  
$W$s were again forced to decay leptonically to enhance statistics.}
\end{table}

\clearpage

\begin{table}[h!]\centering
\scriptsize
\begin{tabular}{| l | ccc | ccc |}
\hline
& \multicolumn{3}{c|}{\bf Point~(600,\,10)L} &   \multicolumn{3}{c|}{\bf Point~(600,\,10)R} \\
\hline
         & SUSY & XQ-SDM & XQ-VDM & SUSY & XQ-SDM & XQ-VDM \\
\hline
efficiency & 0.00314 & 0.00334& 0.00323& 0.00276 & 0.00285 & 0.00286\\
excl. XS [pb] & 0.0470 &  0.0443 & 0.0455 & 0.0535 & 0.0520 & 0.0518\\
mass limit/SUSY XS & 547 & 552 & 550 & 537 & 539 & 540\\
mass limit/XQ XS & 717 & 723 & 720 & 705 & 707 & 708\\
\hline
$1-{\rm CLs}$ &0.79  & 1 & 1 & 0.74 & 1 & 1 \\
\hline
\end{tabular}\\[4mm]
\begin{tabular}{| l | ccc | ccc |}
\hline
& \multicolumn{3}{c|}{\bf Point~(600,\,300)L} &   \multicolumn{3}{c|}{\bf Point~(600,\,300)R} \\
\hline
         & SUSY & XQ-SDM & XQ-VDM & SUSY & XQ-SDM & XQ-VDM \\
\hline
efficiency & 0.00134 & 0.001425 & 0.00138 & 0.00111 & 0.00118 & 0.00100 \\
excl. XS [pb] & 0.109 & 0.104 & 0.108 & 0.133 & 0.125 & 0.148\\
mass limit/SUSY XS & 484 & 487 & 484 & 469 & 473 & 462\\
mass limit/XQ XS & 638 & 642 & 639 & 620 & 626 & 611\\
\hline
$1-{\rm CLs}$ &0.49  & 1 & 1 & 0.43 & 1 & 1 \\
\hline
\end{tabular}
\caption{\label{tab:lept-eff-limits-atlas} Efficiencies, cross-sections excluded at 95\%~CL, corresponding extrapolated top-partner mass limits in GeV, and CLs exclusion value for the ATLAS stop search in the 2-lepton channel, derived with {\sc CheckMATE}. All numbers correspond to the most sensitive signal region, SR5.}
\end{table}

%\clearpage
%-----------------------------------------------------------------------------
\subsection{Gluino/squark search in the 2--6 jets final state \label{sec:gluinosquark}}
\label{sec:genglusq}
%-----------------------------------------------------------------------------

For completeness, we also include a generic SUSY search (nominally for squarks and gluinos) in final states containing 
high-$p_T$ jets, missing transverse momentum and no electrons or muons in our analysis. Concretely, we here consider the 
ATLAS analysis~\cite{Aad:2014wea} via the {\sc MadAnalysis}\,5 recast code~\cite{MA5-ATLAS-multijet-1405}. (A {\sc CheckMATE} implementation of the same analysis was done in \cite{Cao:2015ara} and will be used in Appendix~A). 
Our original purpose was to compare the performance of the hadronic stop analysis  to that of a multi-jet analysis which was not optimized for the $t\bar t+\MET$ signature. But, as we will see, the effective mass $M_{\rm eff}$ variable employed in the 
generic gluino/squark search offers a useful complementary probe. 

Regarding the signal selection, the ATLAS analysis~\cite{Aad:2014wea} comprises 15 inclusive SRs characterized by increasing minimum jet multiplicity, $N_j$, from two to six jets. 
Hard cuts are placed on missing energy and the $p_T$ of the two leading jets: $\MET>160$~GeV, $p_T(j_1)>130$~GeV and $p_T(j_2)>60$~GeV.  
For the other jets, $p_T>60$~or 40~GeV is required depending on the SR. 
In all cases, events are discarded if they contain electrons or muons with $p_T>10$~GeV. 
Depending on $N_j$, additional requirements are placed on the minimum azimuthal separation between any of the jets and the $\MET$, $\Delta\phi({\rm jet},\MET)$, as well as on $\MET/\sqrt{H_T}$ or $\MET/M_{\rm eff}(N_j)$. 
Finally, a cut is placed on $M_{\rm eff}({\rm incl.})$, which sums over all jets with $p_T>40$~GeV and $\MET$.
A cut-flow example is shown in Table~\ref{tab:cutflow-gluino-squark} for Point (600,10)R for a SR with 4 jets (SR {\tt 4jl}). 
Note that, starting from 200K events, we end up with about 15\% (11\%) more SUSY than XQ-SDM (XQ-VDM) events in this SR. 
The reason for this is that the cuts on $p_T(j)$ and $M_{\rm eff}$ remove somewhat more XQ than SUSY events, as expected from the distributions in Fig.~\ref{fig:GenLevelDist1}. 

Table~\ref{tab:limits-gluino-squark} summarises the total efficiencies in the most important SRs of this analysis together with the cross-sections excluded at 95\%~CL and the corresponding estimated top-partner mass limits for all four benchmark scenarios. 
We observe about 20\% difference in the excluded cross-sections between SUSY and XQ interpretations. 
However, the mass limits derived from the excluded cross-sections are not reliable because for this search the total efficiencies strongly depend on the top-partner mass. 
As we will see in the next section, while this analysis does provide a limit on $T\bar T$ production because of the larger cross-section, it is not sensitive to $\st_1^{}\st_1^*$ production.

\begin{table}[t!]\centering 
\scriptsize
\begin{tabular}{lccc}
\hline
         & SUSY & XQ-SDM & XQ-VDM \\
\hline        
Initial no.\ of events & 200000 & 200000 & 200000 \\ 
$\MET > 160$~GeV & 158489 (-20.76\%) & 158497 (-20.75\%) & 159683 (-20.16\%) \\
$N_j > 1$ & 150908 (-4.78\%) & 150121 (-5.28\%) & 151311 (-5.24\%) \\ 
lepton veto & 100139 (-33.64\%) & 100462 (-33.08\%) & 101404 (-32.98\%) \\ 
$p_T(j_1)>130$~GeV & 62585 (-37.50\%) & 58754 (-41.52\%) & 59482 (-41.34\%) \\ 
$p_T(j_2)>60$~GeV & 62045 (-0.86\%) & 58188 (-0.96\%) & 58886 (-1.00\%) \\ 
$p_T(j_3)>60$~GeV & 56729 (-8.57\%) & 52649 (-9.52\%) & 53312 (-9.47\%) \\ 
$p_T(j_4)>60$~GeV & 39150 (-30.99\%) & 34856 (-33.80\%) & 35258 (-33.86\%) \\ 
$\Delta\phi(j_1),\MET) > 0.4$ & 38811 (-0.87\%) & 34616 (-0.69\%) & 35000 (-0.73\%) \\ 
$\Delta\phi(j_2),\MET) > 0.4$ & 37199 (-4.15\%) & 33304 (-3.79\%) & 33635 (-3.90\%) \\ 
$\Delta\phi(j_3),\MET) > 0.4$ & 35447 (-4.71\%) & 31870 (-4.31\%) & 32211 (-4.23\%) \\ 
$\Delta\phi(j_4),\MET) > 0.2$ & 34535 (-2.57\%) & 31064 (-2.53\%) & 31435 (-2.41\%) \\ 
$\MET/\sqrt{H_T}>10$ & 25451 (-26.30\%) & 23522 (-24.28\%) & 24004 (-23.64\%) \\ 
$M_{\rm eff}({\rm incl.})>1$~TeV & 17695 (-30.47\%) & 15062 (-35.97\%) & 15714 (-34.54\%) \\ 
\hline
\end{tabular}
\caption{\label{tab:cutflow-gluino-squark} Cut-flow for the {\tt 4jl} SR of the ATLAS gluino and squark search in the 2--6 jets channel for Point~(600,\,10)R, derived with the {\sc MadAnalysis}\,5 recast code~\cite{MA5-ATLAS-multijet-1405}.}
\end{table}

\begin{table}[t!]\centering
\scriptsize
\begin{tabular}{| l | ccc | ccc |}
\hline
& \multicolumn{3}{c|}{\bf Point~(600,\,10)L} &   \multicolumn{3}{c|}{\bf Point~(600,\,10)R} \\
\hline
         & SUSY & XQ-SDM & XQ-VDM & SUSY & XQ-SDM & XQ-VDM \\
\hline
efficiency &  0.08898 & 0.07454 & 0.07752  &  0.08847 & 0.07531  &  0.07857  \\
excl. XS [pb] & 0.0535 & 0.0639 & 0.0612 &  0.0538 & 0.0631 & 0.0605   \\
mass limit/SUSY XS  &  537 & 523 & 527  &   537 & 524 & 528 \\
mass limit/XQ XS      &  705 & 688 & 692   &   704 & 689 & 693   \\
\hline
$1-{\rm CLs}$ &0.65  & 1 & 1 & 0.66 & 1 & 1 \\
\hline
\end{tabular}\\[4mm]
\begin{tabular}{| l | ccc | ccc |}
\hline
& \multicolumn{3}{c|}{\bf Point~(600,\,300)L} &   \multicolumn{3}{c|}{\bf Point~(600,\,300)R} \\
\hline
         & SUSY & XQ-SDM & XQ-VDM & SUSY & XQ-SDM & XQ-VDM \\
\hline
efficiency &  0.05183 & 0.04242 & 0.04159  &  0.05231 & 0.04281  &  0.04020 \\
excl. XS [pb] &  0.257 & 0.313 & 0.320 &   0.254 & 0.311 & 0.330   \\
mass limit/SUSY XS  &  424 & 410 & 409  &   424 & 411 & 407  \\
mass limit/XQ XS      &  563 & 547 & 545  &  564 & 547 & 542   \\
\hline
$1-{\rm CLs}$ &0.13  & 0.67 & 0.66 & 0.13 & 0.68 & 0.65 \\
\hline
\end{tabular}
\caption{\label{tab:limits-gluino-squark} Efficiencies, cross-sections excluded at 95\%~CL and corresponding extrapolated top-partner mass limits in GeV for the ATLAS gluino and squark search in the 2--6 jets channel, derived with the {\sc MadAnalysis}\,5 recast code~\cite{MA5-ATLAS-multijet-1405}. The last entry is the CLs exclusion value. The most sensitive SR is 4jl for the (600,\,10) mass combination and 4jlm for the (600,\,300) mass combination. Note that for this search the efficiencies strongly depend on the top-partner mass, so the extrapolation of the mass limit is unreliable; this is to large extent due to the cut on $M_{\rm eff}$.}
\end{table}

%\clearpage
%==============================================================================
\section{Results in the  top-partner versus DM mass plane}
\label{sec:bounds8TeV}
%==============================================================================

Having analysed the differences, or lack thereof, in the cut efficiencies of the experimental analyses for our four benchmark points, we next perform a scan in the plane of top-partner versus DM mass to derive the 95\% CL exclusion lines. 
For definiteness, we keep the couplings fixed to the same values as for the (600,\,10)L and (600,\,10)R benchmark points. 

\begin{figure}[t!]\centering 
\includegraphics[width=0.48\textwidth]{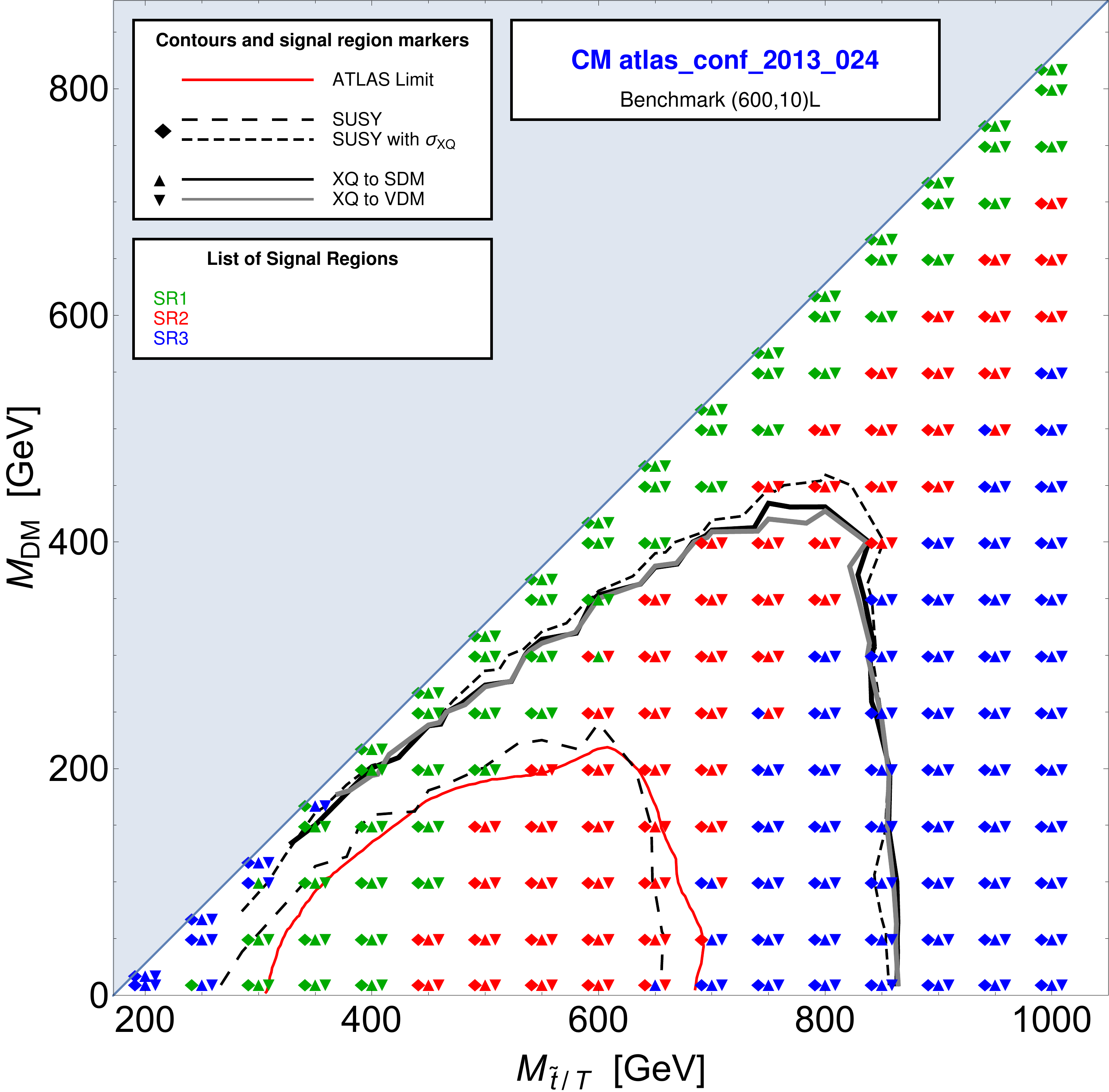}\quad%
\includegraphics[width=0.48\textwidth]{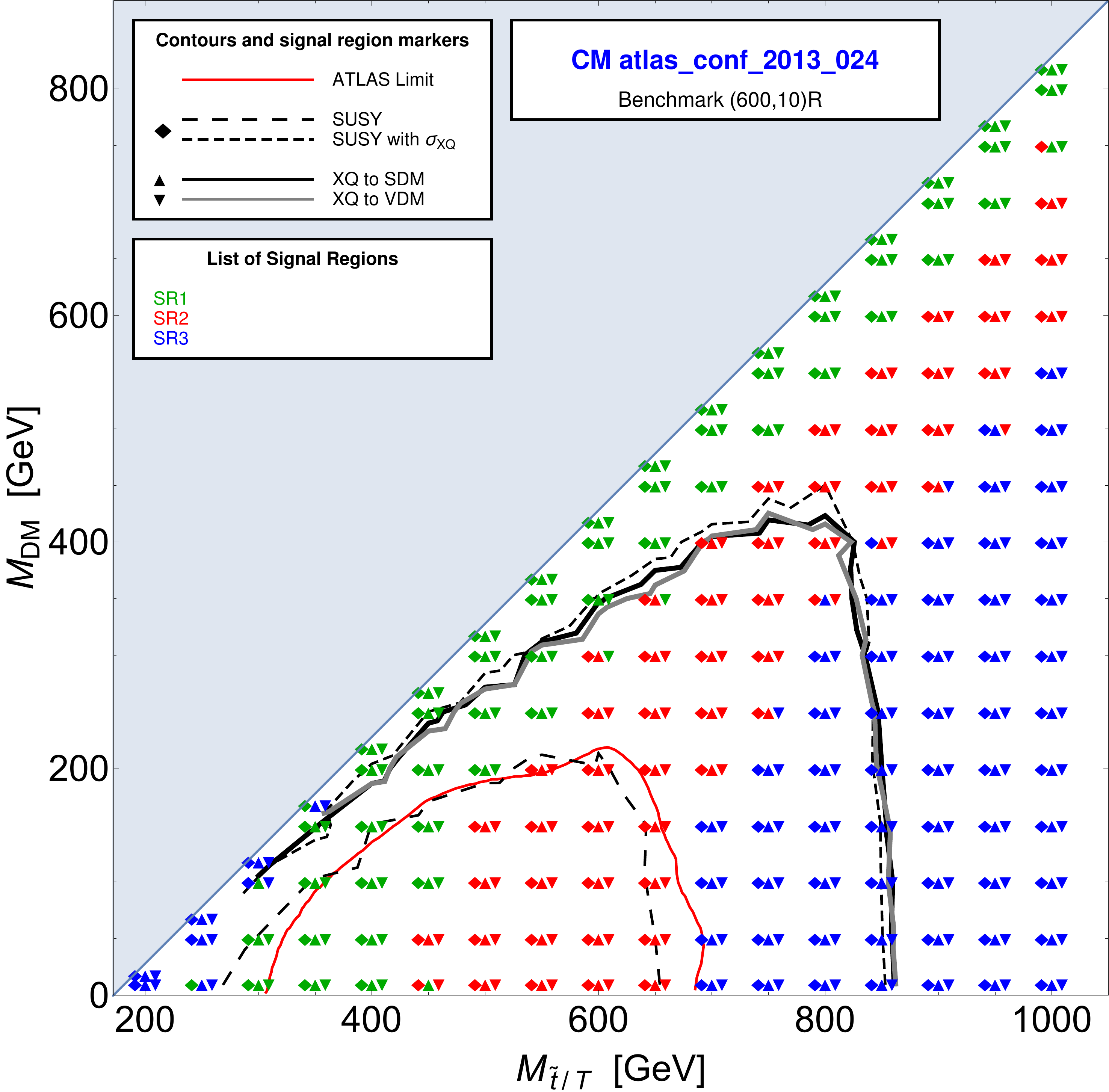}\\[3mm]
\includegraphics[width=0.48\textwidth]{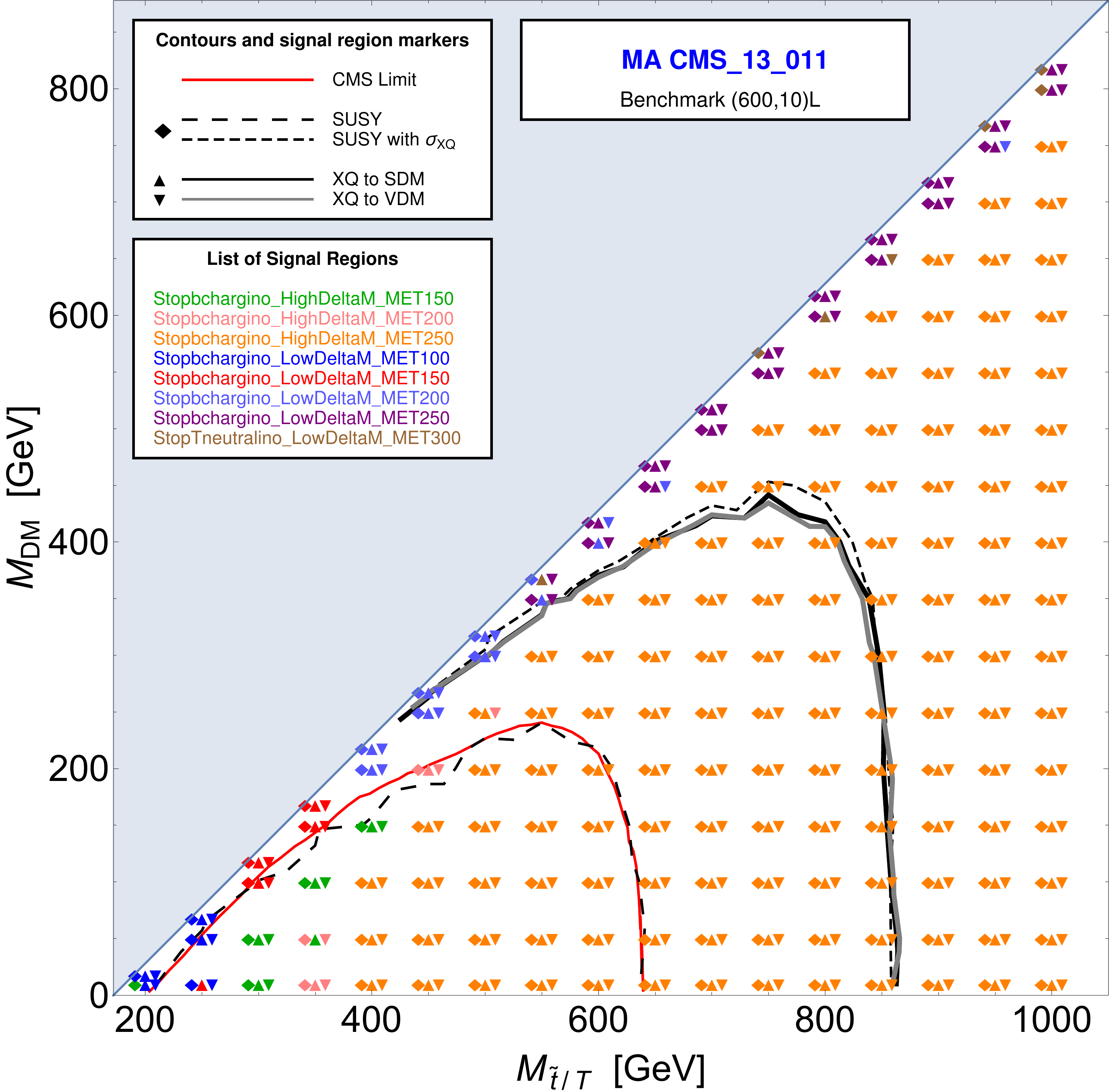}\quad%
\includegraphics[width=0.48\textwidth]{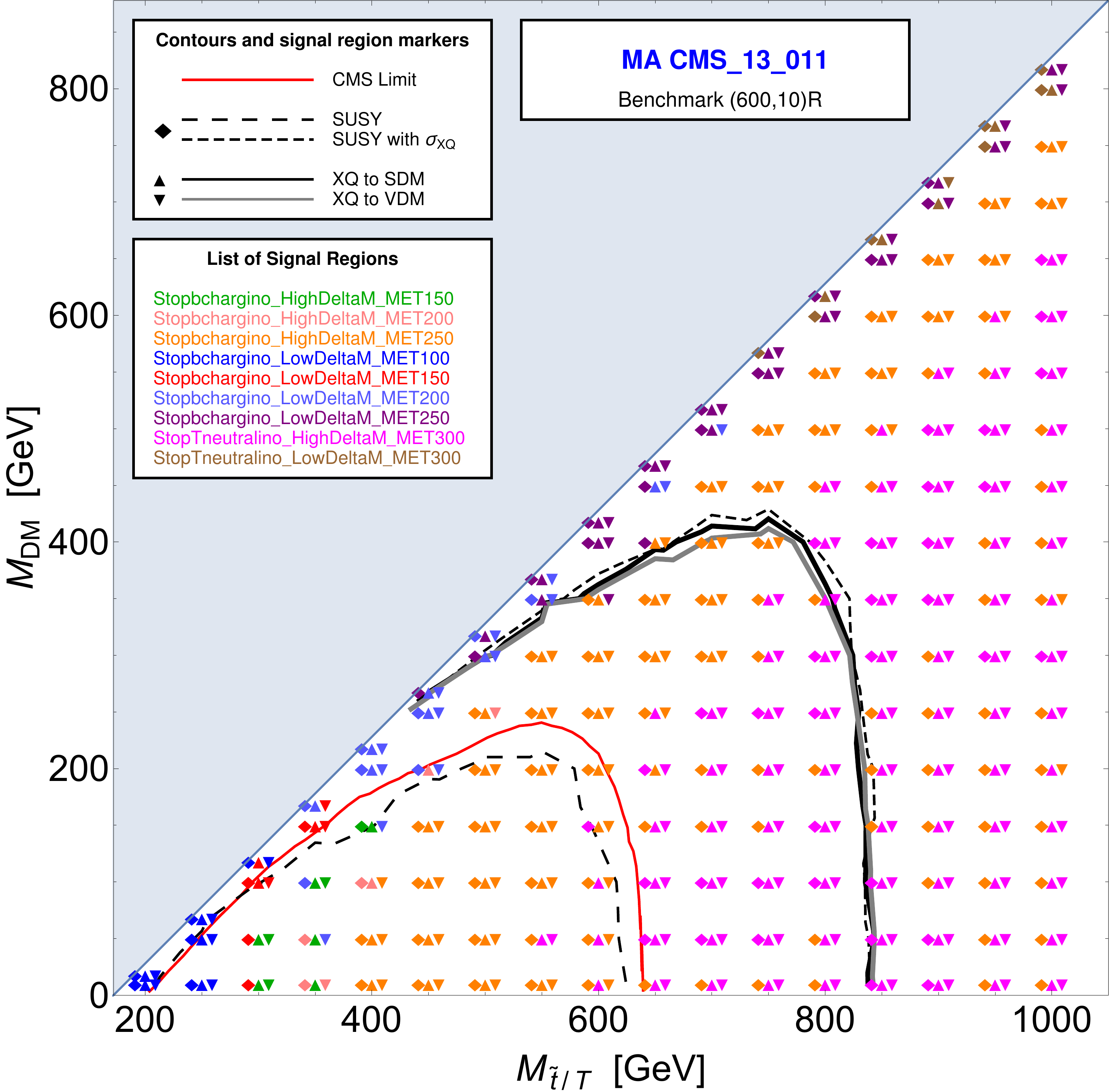}\\[3mm]
\includegraphics[width=0.48\textwidth]{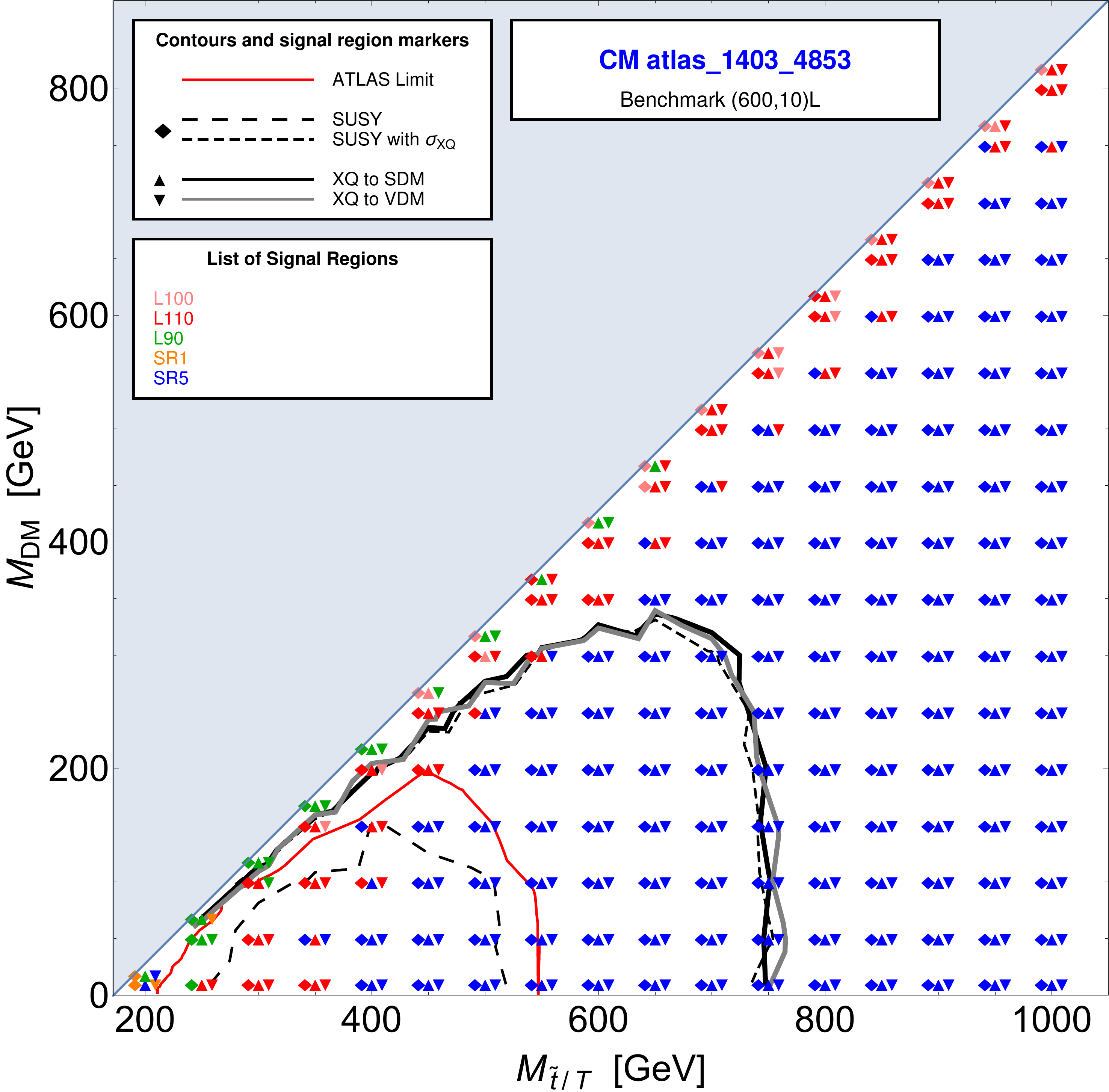}\quad%
\includegraphics[width=0.48\textwidth]{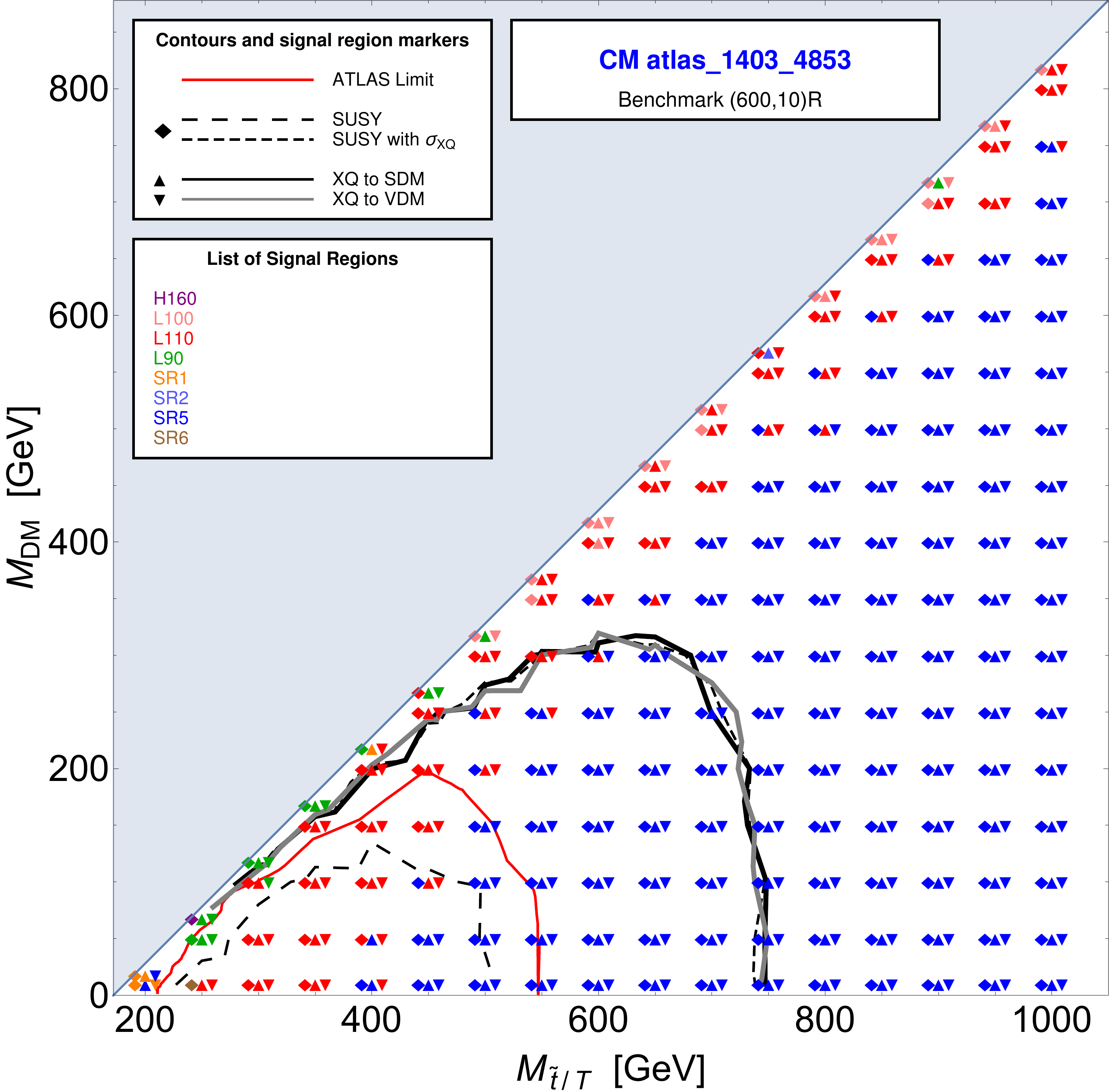}%
\caption{\label{fig:contours1} Comparisons of constraints in the top-partner versus DM mass plane  
for the fully hadronic stop search from ATLAS recast with {\sc CheckMATE} (top), the 1-lepton stop search from CMS recast with {\sc MadAnalysis}\,5 (middle), and the 2-lepton stop search from ATLAS recast with {\sc CheckMATE} (bottom). 
See text for details. 
%The various lines indicate the regions excluded at 8 TeV for the SUSY and XQ cases, and for the case where the SUSY efficiencies are applied to the XQ cross-sections. The plots also contain the information which SRs are the most sensitive ones for each point of the scan. For reference, the official ATLAS/CMS exclusion lines are also shown (full red lines).
}\end{figure}

Figure~\ref{fig:contours1} presents the results for the ATLAS fully hadronic stop search implemented in {\sc CheckMATE} (top row), the CMS 1-lepton stop search recast with {\sc MadAnalysis}\,5 (middle row) and the ATLAS stop search in the 2-lepton final state recast with {\sc CheckMATE} (bottom row). 
The left panels are for the couplings of Point (600,\,10)L, the right panels for the couplings of Point (600,\,10)R, see Table~\ref{tab:BPlist}. 
Shown are the 95\%~CL exclusion lines obtained from SUSY, XQ-SDM and XQ-VDM event simulation (dashed black, full black and full grey lines, respectively), as well as the exclusion lines obtained from rescaling SUSY efficiencies with XQ cross-sections (dotted black line). For each bin, the most sensitive SR used for the limit setting in the SUSY, XQ-SDM and XQ-VDM case is indicated by a coloured symbol as shown in the plot legends.
For reference, the official ATLAS/CMS exclusion lines are also shown as full red lines.

For the CMS 1-lepton search, our  exclusion line for left stops agrees remarkably well with the official CMS line (from the cut-based analysis). This is somewhat accidental, as {\it i)} the official CMS limit is for for unpolarised stops, and {\it ii)} in our simulation the limit is mostly obtained from a SR optimised for decays to bottom and chargino, not from one optimised for decays to top and neutralino. 
On the other hand, the fairly large discrepancy for the ATLAS 2-lepton search is explained by the fact that the official exclusion curve was obtained using an MVA not available in {\sc CheckMATE}.

We see that over most of the mass plane, the best SR is the same for SUSY, XQ-SDM and XQ-VDM. 
(For the points where they are different, the sensitivities of the best and 2nd best SRs are actually quite similar.) 
The main conclusions which can be inferred from the plots are the following: 

\begin{enumerate}
\item There are no significant differences between the XQ scenarios where the top partner decays to scalar or vector DM. This is expected because in the narrow-width approximation the process is largely dominated by the resonant contribution, the cross-section of which can be factorised into production cross-section times branching ratios. Since in our framework the branching ratios are 100\% in the $t+{\rm DM}$ channel, there are no relevant differences between different DM hypotheses. 
\item The contours obtained by rescaling the SUSY efficiencies with the XQ cross-sections coincide quite well with the ``true'' XQ exclusion lines obtained by simulating XQ events. 
This means, efficiency maps or cross-section upper limit maps for the stop--neutralino simplified model can safely be applied to the XQ case under consideration in this paper. It would thus be of advantage if the official maps by ATLAS and CMS extended to high enough masses to cover the 95\% CL reach for fermionic top partners, which is currently not the case.  
\end{enumerate}

\begin{figure}[t!]\centering 
\includegraphics[width=0.48\textwidth]{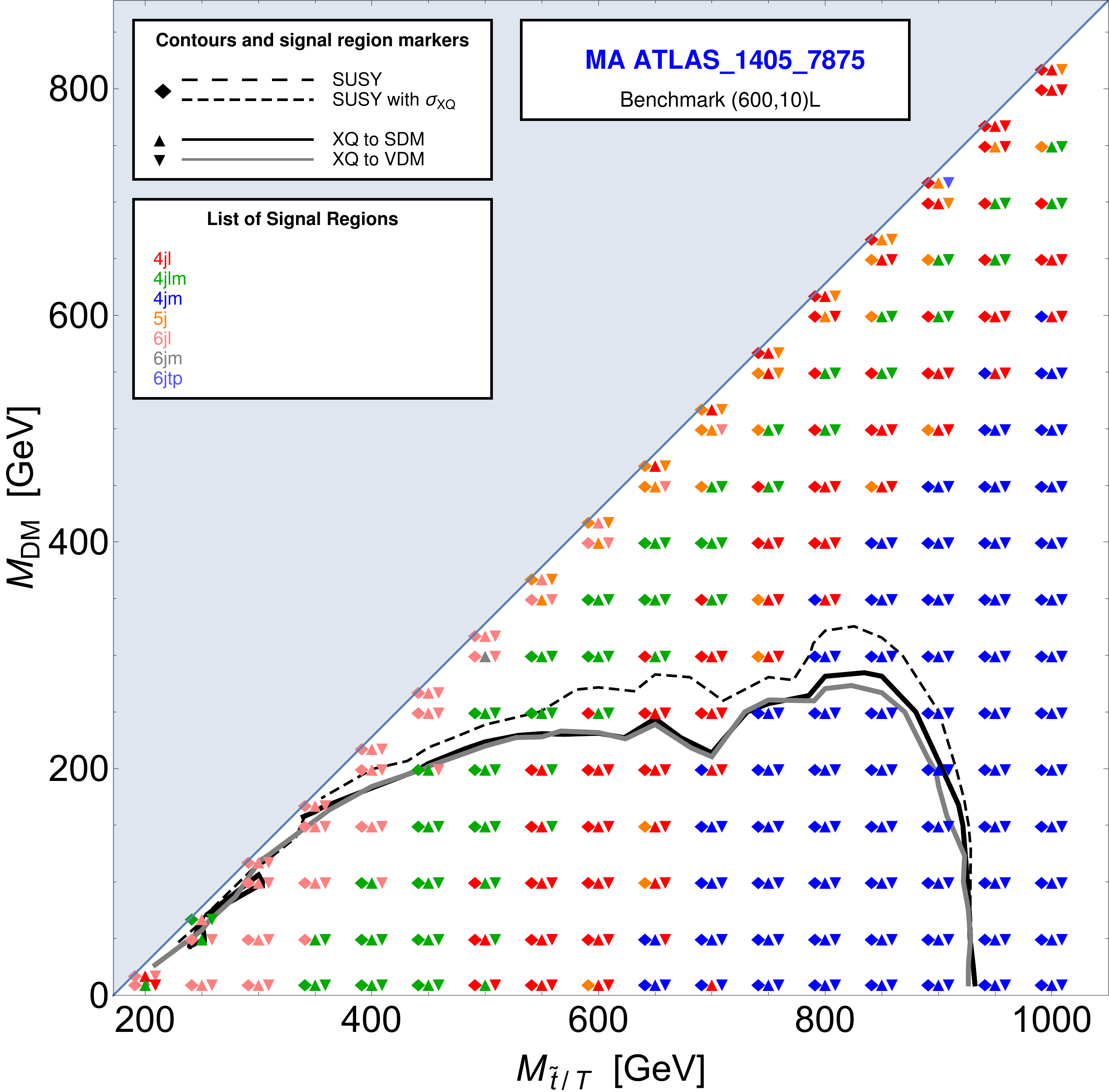}\quad%
\includegraphics[width=0.48\textwidth]{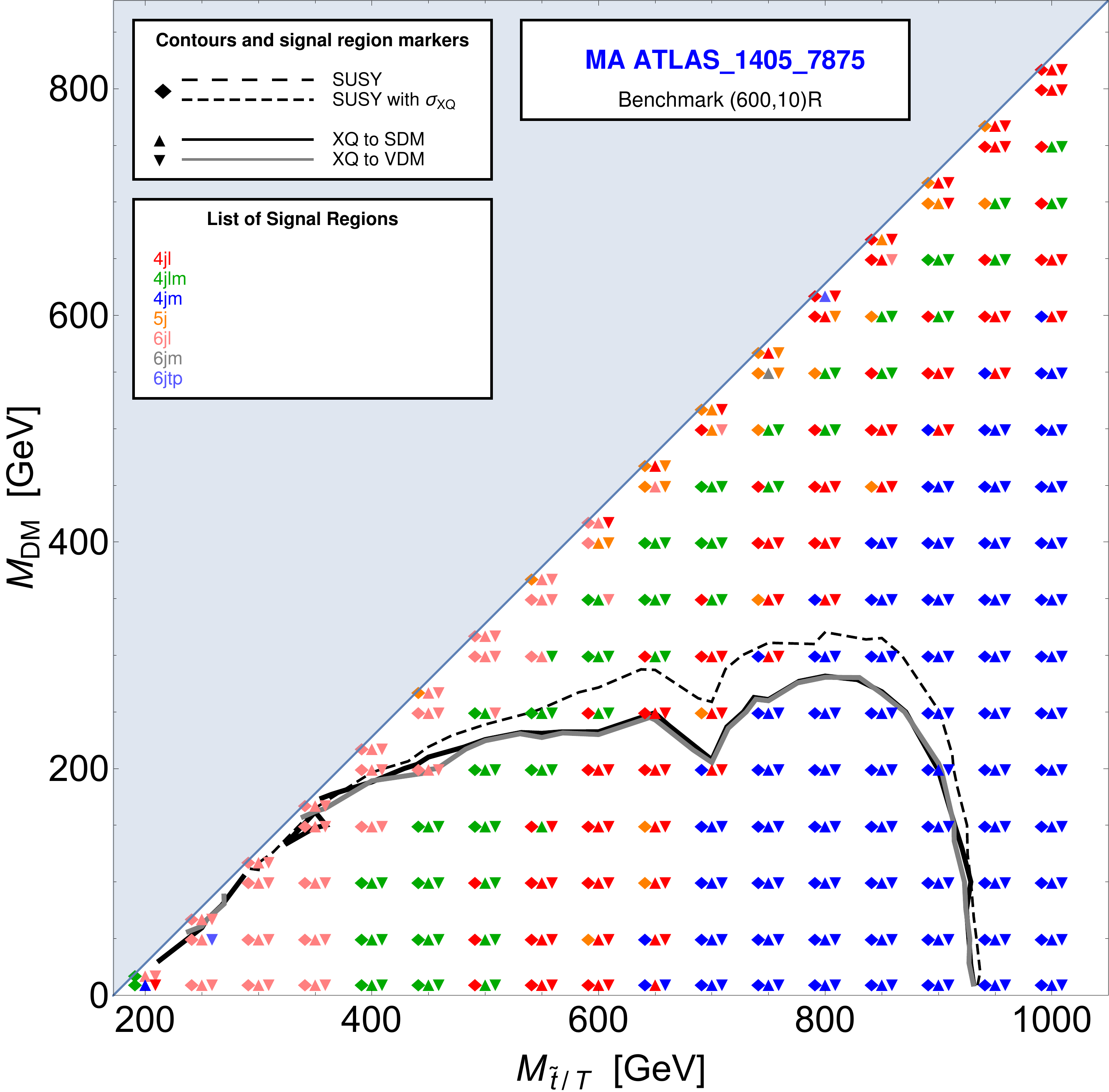}
\caption{\label{fig:contours2} Comparison of constraints in the top-partner versus DM mass plane based on the {\sc MadAnalysis}\,5 recast code for the ATLAS gluino/squark search with 2--6 jets. As in Fig.~\ref{fig:contours1}, 
the various lines indicate the regions excluded at 8 TeV for the SUSY and XQ cases, and for the case where the SUSY efficiencies are applied to the XQ cross-sections. The plots also contain the information which SRs are the most sensitive ones for each point of the scan. Note that no stop--neutralino mass limit is obtained from this analysis. 
}\end{figure}

\noindent 
The situation is different for the generic gluino/squark search in the multi-jet + $\ETmiss$ channel shown in Fig.~\ref{fig:contours2}.\footnote{To produce this figure, we have extended the {\sc MadAnalysis}\,5 recast code with the SRs {\tt 2jl}, {\tt 4jm} and {\tt 6jm}, which are not present in the PAD version \cite{MA5-ATLAS-multijet-1405}. We note, however, that these SRs could not be validated, as no cut-flows or kinematic distributions are available for them from ATLAS.}  
Contrary to the estimated stop mass limit of about 400--500~GeV in Table~\ref{tab:limits-gluino-squark},    
in the scan we do not obtain any limit on stops from this analysis. As already mentioned in Section~\ref{sec:gluinosquark}, the reason is that the efficiency of the $M_{\rm eff}$ cut strongly depends on the overall mass scale, rendering the extrapolation of the limit unreliable. This can also be seen from the fact that the most sensitive SR changes more rapidly with the top-partner mass, see the colour code in Fig.~\ref{fig:contours2}. (The {\sc CheckMATE} implementation of the same analysis gives slightly stronger constraints on the SUSY case, excluding the region $m_{\tilde t}\approx 300-400$~GeV and $m_{\tilde\chi^0_1}\lesssim 50$~GeV, see the Appendix.)
Likewise, also the limit for the XQ case derived from the scan differs from the estimated one in Table~\ref{tab:limits-gluino-squark}, although here the effect goes in the opposite direction: the actual limit is stronger than the extrapolated one.
In fact, due to the increased efficiencies at high mass scales, this search can give stronger constraints on the XQ case than the stop searches, extending the limit up to $m_T\approx 900$--950~GeV for $m_{\rm DM}\lesssim 300$~GeV. 
The naive rescaling of SUSY efficiencies with XQ cross-sections (dashed lines) however somewhat overestimates the reach for the XQ scenario. For this kind of analysis it will thus be interesting to produce efficiency maps specifically for the XQ model.

\FloatBarrier
%\clearpage

%==============================================================================
\section{Conclusions}\label{sec:conclusions}
%==============================================================================

We have studied how various analyses targeting $t\bar t +\MET$ signatures, carried out 
by ATLAS and CMS in the context of SUSY searches, perform for models with fermionic top partners. 
Taking a simplified XQ model with one extra $T$ quark and one DM state and comparing it to a 
simplified stop--neutralino model, we found that given the same kinematical configuration, SUSY and XQ 
efficiencies are very similar.  
The situation is different for generic multi-jet + $\MET$ searches targeting light-flavour squark and gluino production: 
here we found larger efficiencies for the SUSY than for the XQ case.  

Putting everything together, we conclude that cross-section upper limit maps and efficiency maps obtained for stop simplified models in stop searches can also be applied to analogous models with fermionic top partners and a DM candidate, provided the narrow-width approximation applies. An exception may be the region of very small mass differences, where uncertainties in the total cut efficiencies become sizeable, though this does not influence much the actual limit.\footnote{However, this region could become important for scenarios in which multiple degenerate or nearly-degenerate top-partners occur, as in this case the cross-section might be enhanced by interference effects. Separate efficiency maps for the scalar or fermionic top partners would therefore be useful in this regime.} 
To fully exploit the applicability to different top partner models, we encourage the experimental collaborations to present their 
cross-section upper limit and efficiency maps for a wide enough mass range, covering not only the reach for stops but also the reach for fermionic top partners. 
For the generic multi-jet + $\MET$ searches, on the other hand, it would be worthwhile to have efficiency maps specifically for the XQ model. 
As a service to the reader and potential user of our work, we provide the efficiency maps which we derived with {\sc CheckMATE} and {\sc MadAnalysis\,5} as auxiliary material~\cite{EfficiencyMaps}. 
The numbers of expected background and observed events from the experimental analyses, needed for the statistical interpretation, are summarized in Appendix~B.

\begin{figure}[t!]\centering 
\includegraphics[width=0.66\textwidth]{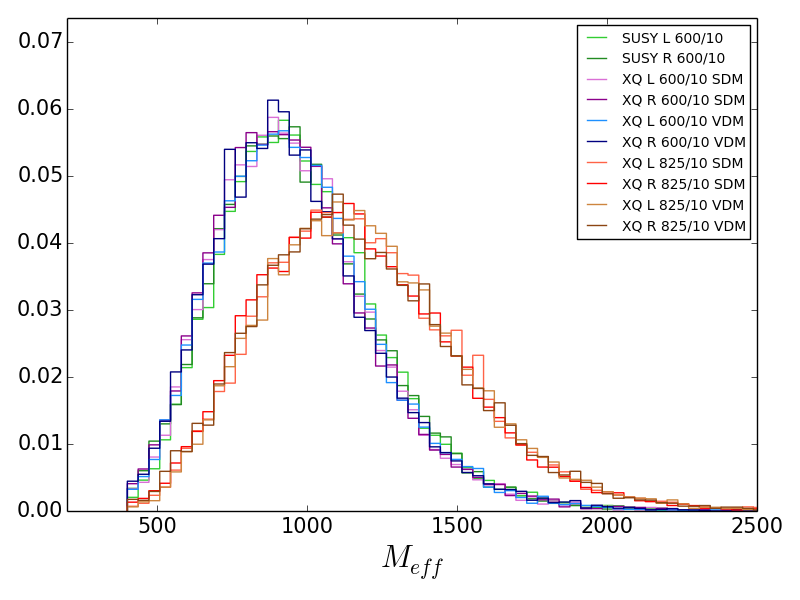}\
\caption{\label{fig:meff} Comparison of the $M_{\rm eff}$ distributions for SUSY and XQ scenarios, after preselection cuts of 
the CMS 1-lepton stop search~\cite{Chatrchyan:2013xna}. Here, $M_{\rm eff}$ is computed as $\sum p_T({\rm jets})+p_T(l)+\MET$. The green, violet and blue histograms are for the default (600,\,10) benchmark points, 
while the orange and brown histograms show XQ scenarios that would give 
roughly the same visible cross-sections as the (600,\,10) SUSY cases.
}\end{figure}

The similarity of SUSY and XQ efficiencies also means that, should a signal be observed in $t\bar t +\MET$ events, 
it is not immediately obvious whether it comes from scalar or fermionic top partners. Since the production cross-section (assumed here to be pure QCD) is significantly larger for fermionic than for scalar top partners, one way of discrimination may be to correlate the effective mass scale, $M_{\rm eff}$, or the effective transverse mass \cite{Cabrera:2012cj}, with the observed number of events, see Fig.~\ref{fig:meff} for an illustrative example. (This was also observed in \cite{Datta:2005vx}. However, as pointed out in \cite{Cheng:2005as}, for small XQ--DM mass splittings the decay products become softer and the discrimination from the SUSY case by cross-section and $M_{\rm eff}$ is lost.)
Moreover, in the case of fermionic top partners, a corroborating signal may show up in generic gluino/squark searches, which have much less sensitivity to scalar top partners.  Finally, the distinction between the two scenarios may be refined by considering special kinematic distributions as discussed in \cite{Han:2008gy,Smillie:2005ar,Datta:2005zs}.

%\clearpage
%==============================================================================
\acknowledgments
%==============================================================================

We thank Daniel Schmeier and Jamie Tattersall for help with {\sc CheckMATE}, and Benjamin Fuks and Dipan Sengupta for help with {\sc MadAnalysis}\,5.

This research was supported in part by the ``Investissements d'avenir, Labex ENIGMASS'', the ANR project DMASTROLHC grant no.\ ANR-12-BS05-0006 and the Theory-LHC-France initiative of the CNRS (INP/IN2P3).
SK thanks the KITP Santa Barbara, supported by the National Science Foundation under Grant No. NSF PHY11-25915, for hospitality during part of the work.

\clearpage
%==============================================================================
\appendix
%==============================================================================

\section{Additional CheckMATE results}

\begin{figure}[t!]\centering 
\includegraphics[width=0.47\textwidth]{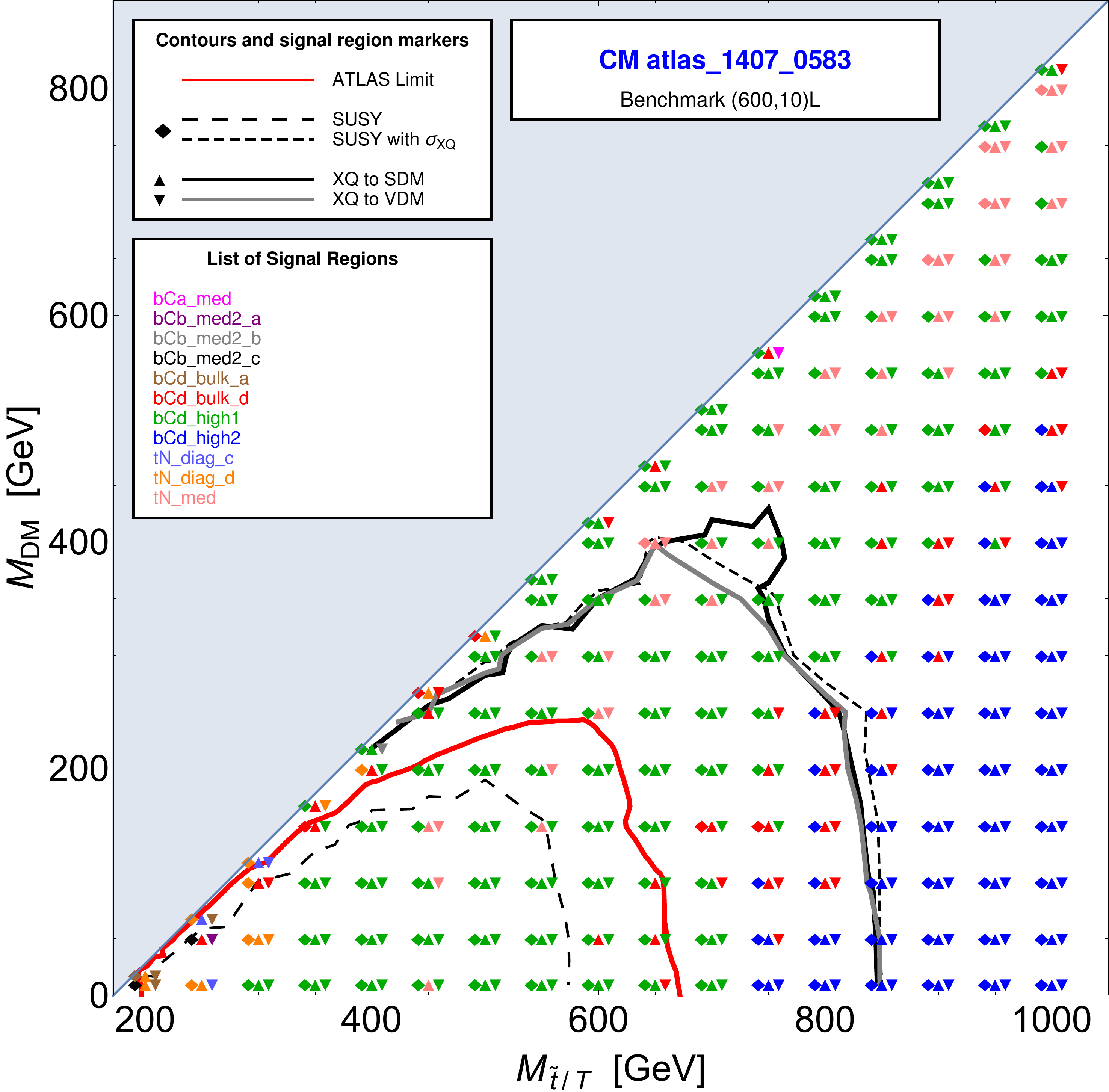}\quad%
\includegraphics[width=0.47\textwidth]{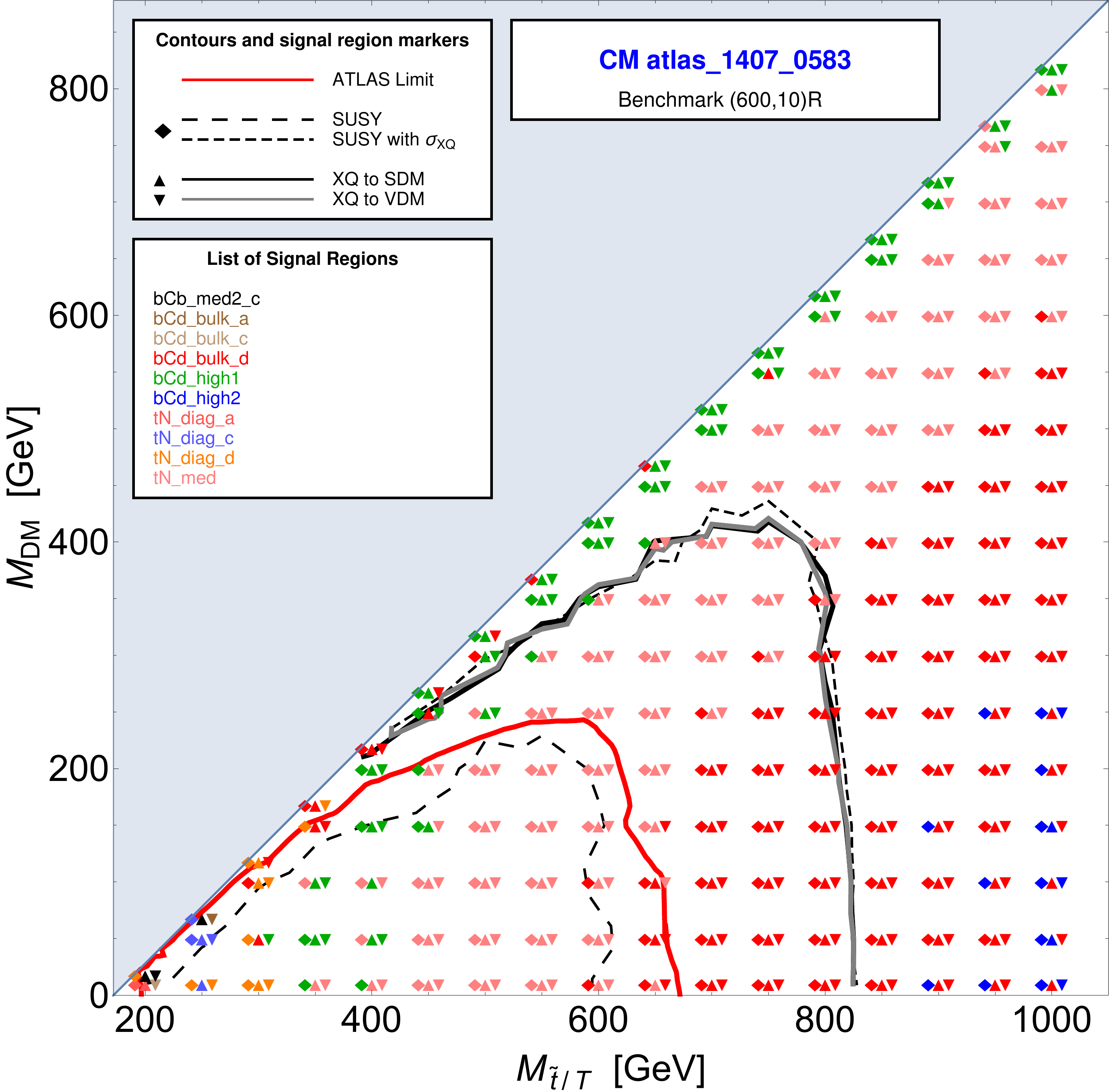}\\[1mm]
\includegraphics[width=0.47\textwidth]{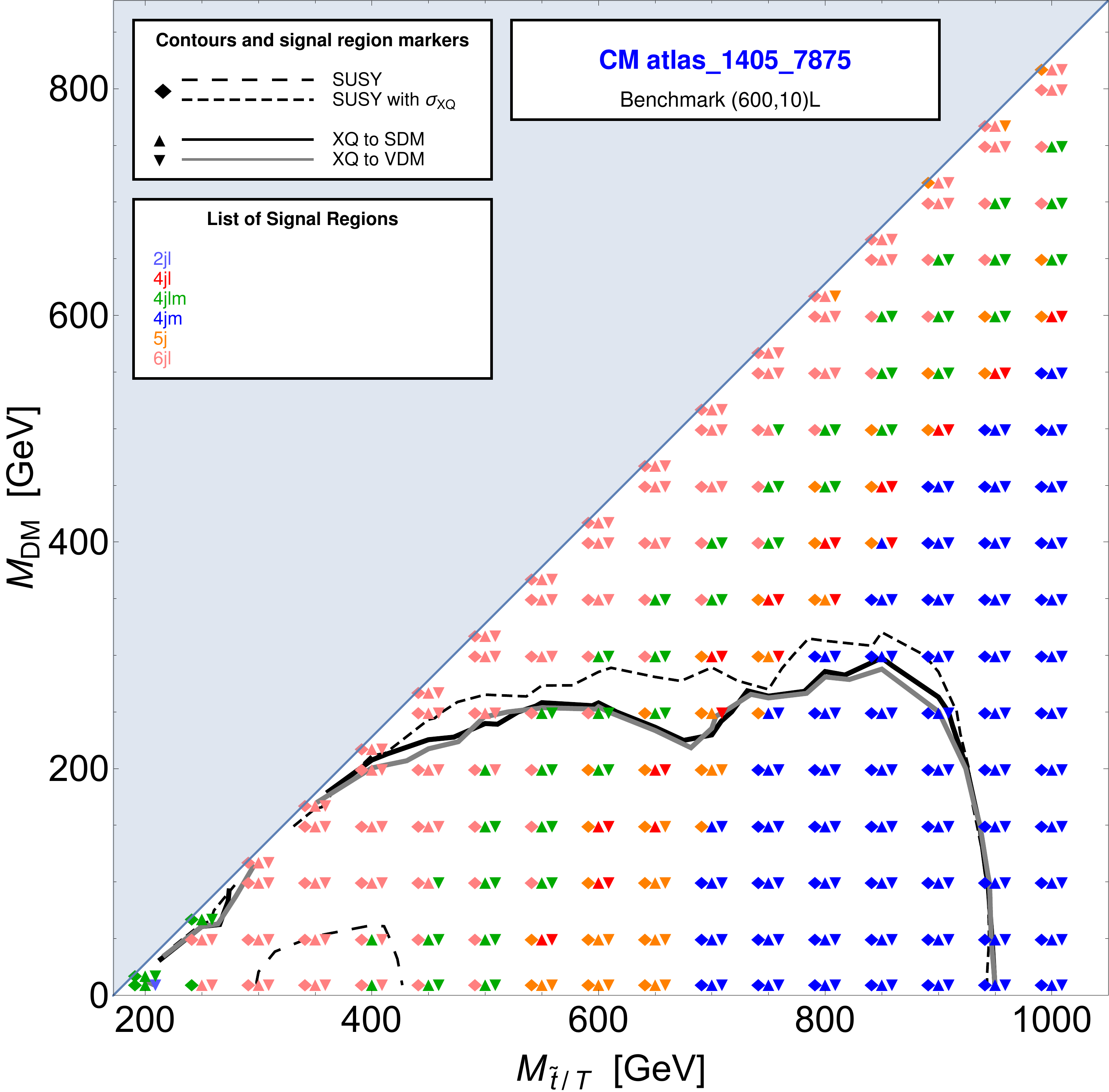}\quad%
\includegraphics[width=0.47\textwidth]{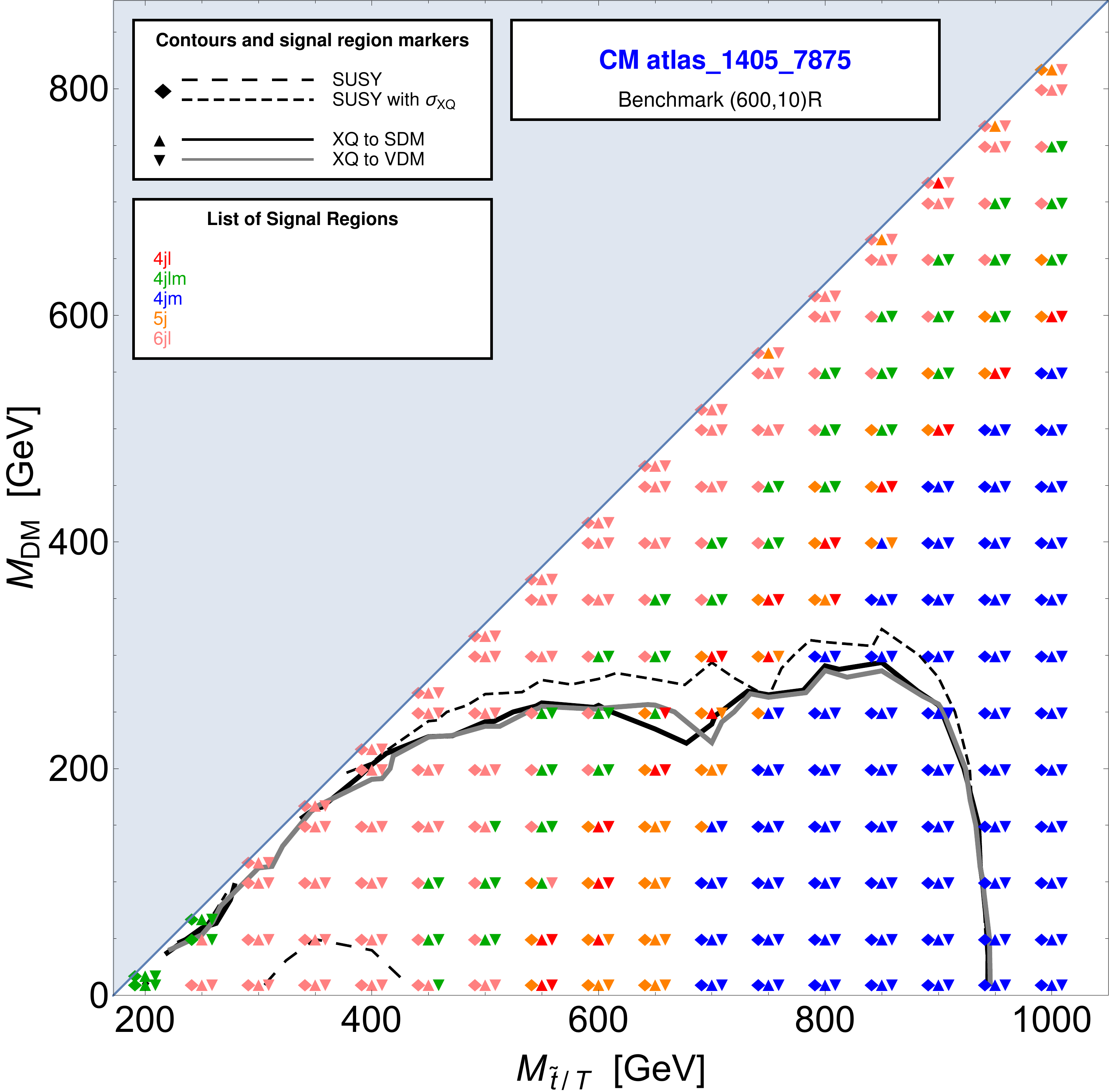}\\[1mm]
\includegraphics[width=0.47\textwidth]{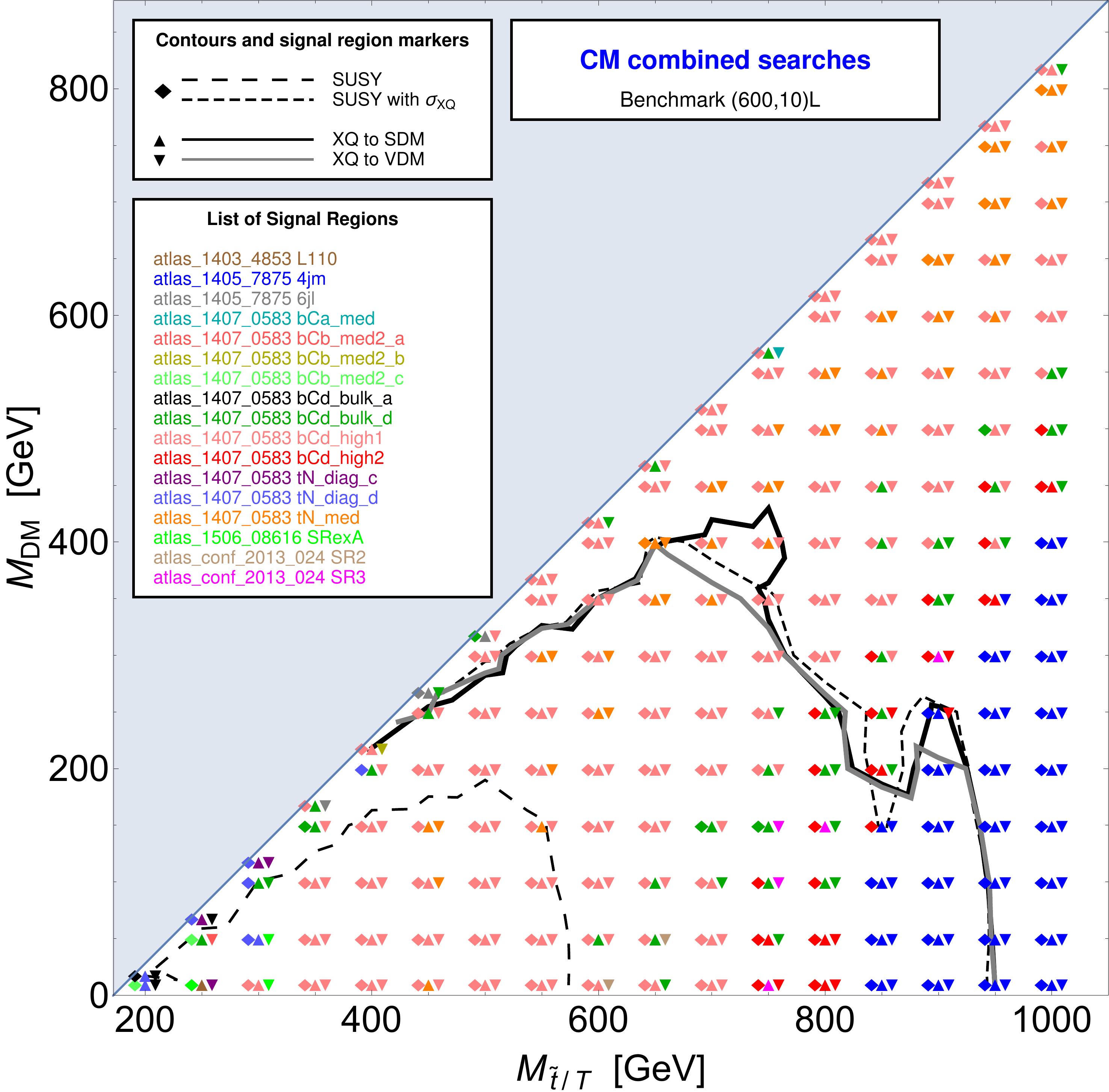}\quad%
\includegraphics[width=0.47\textwidth]{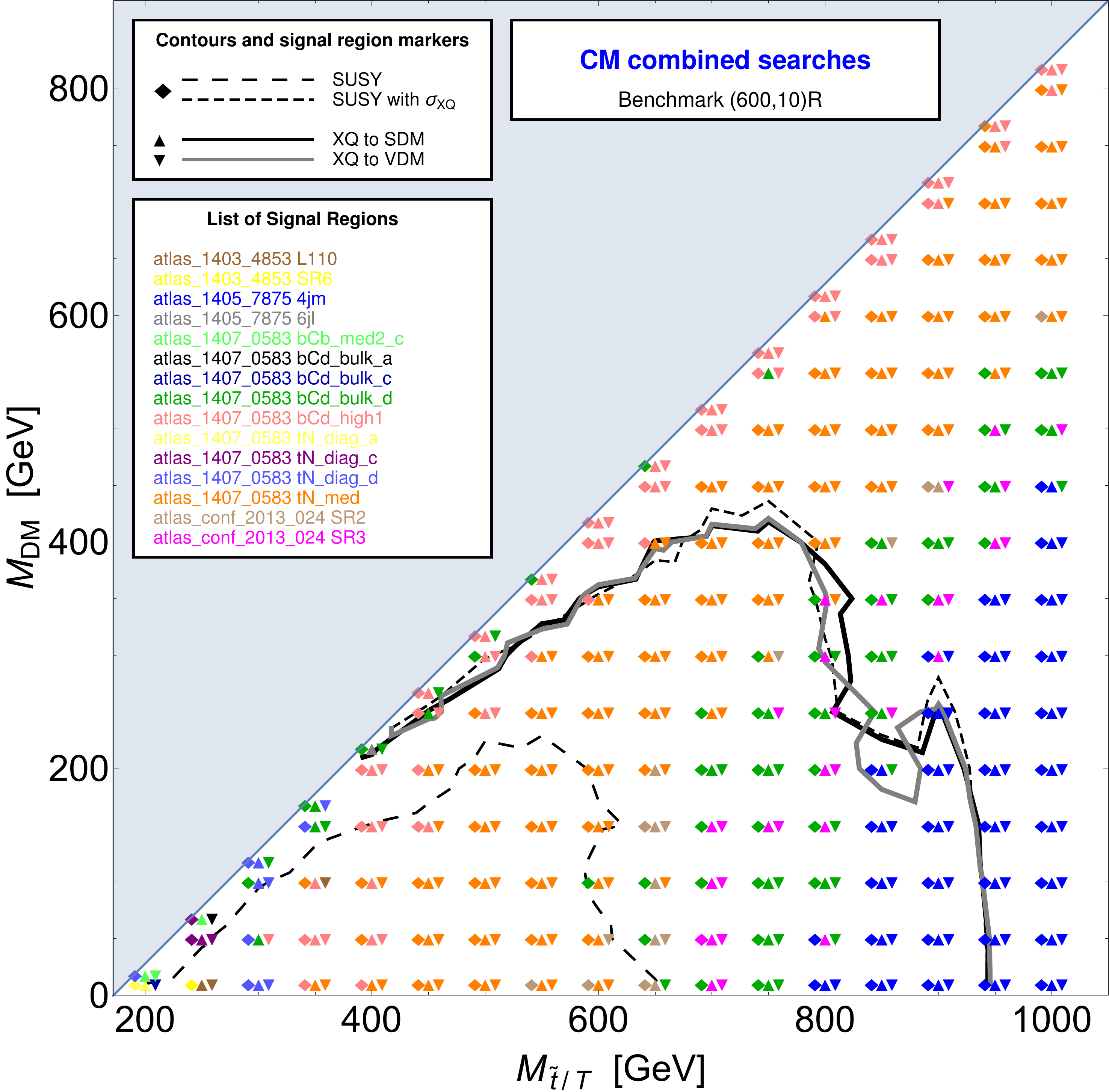}%
\caption{\label{fig:contoursCMall} Additional comparison of constraints in the top-partner versus DM mass plane based on ATLAS analyses implemented in {\sc CheckMATE}: 1-lepton stop search \cite{Aad:2014kra} (top row), generic gluino/squark search \cite{MA5-ATLAS-multijet-1405} (middle row) and combination of all {\sc CheckMATE} ATLAS analyses (bottom row). As before, the left panels are for the couplings of Point (600,\,10)L, the right panels for the couplings of Point (600,\,10)R. 
}\end{figure}

As mentioned in Section~\ref{sec:analyses8tev}, the ATLAS analyses \cite{Aad:2014kra} (1-lepton stop) and \cite{Aad:2014wea} (2--6 jets gluino/squark) are also implemented in {\sc CheckMATE}.  For completeness, we show in Fig.~\ref{fig:contoursCMall} the  {\sc CheckMATE} results for these two analyses together with the constraints obtained when considering all {\sc CheckMATE} ATLAS analyses simultaneously. 

For the 1-lepton stop search from ATLAS, top row in Fig.~\ref{fig:contoursCMall}, we note that the official SUSY limit is less well reproduced than for the corresponding CMS search recast with {\sc MadAnalysis}\,5, cf.\ the middle row of plots in Fig.~\ref{fig:contours1}. This is expected, as the signal region {\tt tN\_boost} of the ATLAS search, which is optimised for high mass scales and boosted tops and is indeed the most sensitive SR for stop masses around 600~GeV, is not implemented in {\sc CheckMATE}. Moreover, there is a larger dependence on the top polarisation, as can be seen from the limit curves but also from  the colour codes identifying the most sensitive SRs. Nonetheless, the resulting limit on XQs is very similar to that obtained from recasting the CMS search with {\sc MadAnalysis}\,5. 
The fact that a stronger limit is obtained for $\tilde t_R^{}$ then for $\tilde t_L^{}$ was also mentioned in the experimental paper, see Fig.~24 in \cite{Aad:2014kra}. 

For the gluino/squark search in the 2--6 jets channel, middle row in Fig.~\ref{fig:contoursCMall}, we observe some differences with respect to the corresponding {\sc MadAnalysis}\,5 results in Fig.~\ref{fig:contours2} in what concerns the best SRs. This can occur when several SRs have comparable sensitivity. The final 95\% CL limit curves for XQs are however very similar in {\sc CheckMATE} and {\sc MadAnalysis\,5}. The main difference is that the {\sc CheckMATE} implementation gives a small exclusion for the SUSY case in the range $m_{\tilde t_1}\approx300$--$400$~GeV and $m_{\tilde\chi_1^0}\lesssim 50$~GeV, while with {\sc MadAnalysis\,5} one obtains only about 80--90\% CL exclusion in this region. 

Running all {\sc CheckMATE} ATLAS analyses simultaneously, one finds that up to top partner masses of about 700~GeV, the 1-lepton stop search \cite{Aad:2014kra} is always more sensitive than the hadronic stop search from the conference note \cite{ATLAS:2013cma}. 
(Although from the top row of plots in Fig.~\ref{fig:contours1} the hadronic analysis seems to give the stronger limit, this comes from the fact that less events were observed in the three SRs of \cite{ATLAS:2013cma} than expected; comparing the expected limits, the search in the 1-lepton channel gives the stronger constraint.) 
It is thus  \cite{Aad:2014kra} which is used for the limit setting in this mass range. Above $m_{T}\approx 700$~GeV, the 
gluino/squark in the 2--6 jets channel \cite{Aad:2014wea} is the most sensitive analysis and used for the limit setting.

\FloatBarrier
\clearpage

\section{Experimental data}

For convenience, we here list in Tables~\ref{tab:expnumbers1}--\ref{tab:expnumbers2} 
the numbers of expected background and numbers of observed events from the experimental analyses used in this paper. 

\begin{table}[h]\centering
\footnotesize
\begin{tabular}{| l | c| c|}
\hline
Signal Region & \# expected events & \# observed events\\
\hline
SR1 & 17.5 $\pm$ 3.2 & 15\\
SR2 & 4.7 $\pm$ 1.5 & 2 \\
SR3 & 2.7 $\pm$ 1.2 & 1\\
\hline
\end{tabular}\\[4mm]
\caption{\label{tab:expnumbers1}  
Results from the fully hadronic stop search from ATLAS~\cite{ATLAS:2013cma}.}
\end{table}

\begin{table}[h]\centering
\footnotesize
\begin{tabular}{| l | c| c|}
\hline
Signal Region & \# expected events & \# observed events\\
\hline
$\tilde t_1\to t+\tilde\chi^0_1$, Low $\Delta M$, $\ETmiss>150$~GeV & 251 $\pm$ 50 & 227 \\
$\tilde t_1\to t+\tilde\chi^0_1$, Low $\Delta M$, $\ETmiss>200$~GeV  & 83 $\pm$ 21 & 69 \\
$\tilde t_1\to t+\tilde\chi^0_1$, Low $\Delta M$, $\ETmiss>250$~GeV  & 31 $\pm$ 8 & 21 \\
$\tilde t_1\to t+\tilde\chi^0_1$, Low $\Delta M$, $\ETmiss>300$~GeV  & 11.5 $\pm$ 3.6 & 9 \\
$\tilde t_1\to t+\tilde\chi^0_1$, High $\Delta M$, $\ETmiss>150$~GeV  & 29 $\pm$ 7 & 23 \\
$\tilde t_1\to t+\tilde\chi^0_1$, High $\Delta M$, $\ETmiss>200$~GeV  & 17 $\pm$ 5 & 11 \\
$\tilde t_1\to t+\tilde\chi^0_1$, High $\Delta M$, $\ETmiss>250$~GeV  & 9.5 $\pm$ 2.8 & 3 \\
$\tilde t_1\to t+\tilde\chi^0_1$, High $\Delta M$, $\ETmiss>300$~GeV  & 4.7 $\pm$ 1.4 & 2 \\
$\tilde t_1\to b+\tilde\chi^+_1$, Low $\Delta M$, $\ETmiss>100$~GeV  & 1662 $\pm$ 203 & 1624 \\
$\tilde t_1\to b+\tilde\chi^+_1$, Low $\Delta M$, $\ETmiss>150$~GeV  & 537 $\pm$ 75 & 487 \\
$\tilde t_1\to b+\tilde\chi^+_1$, Low $\Delta M$, $\ETmiss>200$~GeV  & 180 $\pm$ 28 & 151 \\
$\tilde t_1\to b+\tilde\chi^+_1$, Low $\Delta M$, $\ETmiss>250$~GeV  & 66 $\pm$ 13 & 52 \\
$\tilde t_1\to b+\tilde\chi^+_1$, High $\Delta M$, $\ETmiss>100$~GeV  & 79 $\pm$ 12 & 90 \\
$\tilde t_1\to b+\tilde\chi^+_1$, High $\Delta M$, $\ETmiss>150$~GeV  & 38 $\pm$ 7 & 39 \\
$\tilde t_1\to b+\tilde\chi^+_1$, High $\Delta M$, $\ETmiss>200$~GeV  & 19 $\pm$ 5 & 18 \\
$\tilde t_1\to b+\tilde\chi^+_1$, High $\Delta M$, $\ETmiss>250$~GeV  & 9.9 $\pm$ 2.7 & 5 \\
\hline
\end{tabular}\\[4mm]
\caption{Results from the 1-lepton stop search from CMS~\cite{Chatrchyan:2013xna}.}
\end{table}

\begin{table}[h]\centering
\footnotesize
\begin{tabular}{| l | c| c|}
\hline
Signal Region & \# expected events & \# observed events\\
\hline
tN\_med &   13 $\pm$ 2.2 & 12 \\
tN\_high &  5 $\pm$ 1 & 5 \\
bCa\_low & 6.5 $\pm$   1.4 & 11 \\
bCa\_med &  17 $\pm$    4  & 20 \\
bCb\_med1 & 32 $\pm$    5 & 41 \\
bCb\_high & 9.8 $\pm$   1.6 & 7 \\
bCc\_diag &  470 $\pm$   50 & 493 \\
bCd\_high1 &  11.0 $\pm$  1.5 & 16 \\
bCd\_high2 & 4.4 $\pm$   0.8 & 5 \\
tNbC\_mix &   7.2 $\pm$   1 & 10 \\
tN\_diag\_a & 136 $\pm$   22 & 117 \\
tN\_diag\_b &  152 $\pm$   20 & 163 \\
tN\_diag\_c &  98 $\pm$    13 & 101 \\
tN\_diag\_d & 236 $\pm$   29 & 217 \\
bCb\_med2\_a & 12.1 $\pm$  2.0  & 10 \\
bCb\_med2\_b &  7.4 $\pm$   1.4 & 10 \\
bCb\_med2\_c &  21 $\pm$ 4 & 16 \\
bCb\_med2\_d &   9.1  $\pm$  1.6 & 9 \\
bCd\_bulk\_a &  133  $\pm$  22 & 144 \\
bCd\_bulk\_b &   73 $\pm$    8 & 78 \\
bCd\_bulk\_c &   66  $\pm$   6 & 61 \\
bCd\_bulk\_d &  26.5 $\pm$  2.6 & 29 \\
threeBody\_a &  16.9 $\pm$  2.8 & 12 \\
threeBody\_b &  8.4 $\pm$   2.2 & 8 \\
threeBody\_c &  35  $\pm$   4 & 29 \\
threeBody\_d &   29 $\pm$    5 & 22 \\
\hline
\end{tabular}\\[4mm]
\caption{Results from the 1-lepton stop search from ATLAS~\cite{Aad:2014kra}.}
\end{table}

\begin{table}[h]\centering
\footnotesize
\begin{tabular}{| l | c| c|}
\hline
Signal Region & \# expected events & \# observed events\\
\hline
L90  & 300 $\pm$ 50       & 274 \\
L100 & 5.2 $\pm$ 2.2     & 3 \\
L110 &  9.3 $\pm$ 3.5    &  8 \\
L120  & 19 $\pm$  9       & 18 \\
H160 & 26 $\pm$  6      &  33 \\
SR1 & 270 $\pm$ 40    &   250 \\
SR2 &  3.4  $\pm$ 1.8   &   1 \\
SR3 &  1.3 $\pm$ 0.6  &    2 \\
SR4 &  3.7 $\pm$ 2.7  &    3 \\
SR5 &  0.5 $\pm$ 0.4  &    0 \\
SR6 &  3.8 $\pm$ 1.6  &    3 \\
SR7 &  15 $\pm$  7    &    15 \\
\hline
\end{tabular}\\[4mm]
\caption{Results from 2-lepton stop search from ATLAS~\cite{Aad:2014qaa}.}
\end{table}

\begin{table}[h]\centering
\footnotesize
\begin{tabular}{| l | c| c|}
\hline
Signal Region & \# expected events & \# observed events\\
\hline
2jl & 13000 $\pm$ 1000 & 12315 \\
2jm & 760 $\pm$ 50 & 715 \\
2jt & 125 $\pm$ 10 & 133 \\
3j & 5.0 $\pm$ 1.2 & 7 \\
4jlm & 2120 $\pm$ 110 & 2169 \\
4jl & 630 $\pm$ 50 & 608 \\
4jm & 37 $\pm$ 6 & 24 \\
4jt & 2.5 $\pm$ 1.0 & 0 \\
5j & 126 $\pm$ 13 & 121 \\
6jl & 111 $\pm$ 11 & 121 \\
6jm & 33 $\pm$ 6 & 39 \\
6jt & 5.2 $\pm$ 1.4 & 5 \\
6jtp & 4.9 $\pm$ 1.6 & 6 \\
\hline
\end{tabular}\\[4mm]
\caption{\label{tab:expnumbers2} Results from the generic squark and gluino search from ATLAS~\cite{Aad:2014wea}.}
\end{table}

\clearpage
%==============================================================================
%\bibliography{references}

\providecommand{\href}[2]{#2}\begingroup\raggedright\endgroup

%==============================================================================
\end{document}